\documentclass[
twocolumn,
pra,showpacs,preprintnumbers,amsmath,amssymb]{revtex4}

\usepackage{times}
\usepackage{graphicx}
\usepackage{amssymb}
\usepackage{amsthm}
\usepackage{amsmath}
\usepackage{dsfont}
\usepackage{bm}
\usepackage{mathrsfs}
\usepackage{bbold}

\newcommand*{\even}{{\cal{E}}}
\newcommand*{\odd}{{\cal{O}}}

\begin{document}

\title{Casimir stress inside planar materials}
\author{Itay Griniasty and Ulf Leonhardt}
\affiliation{
\normalsize{
Department of Physics of Complex Systems,
Weizmann Institute of Science, Rehovot 761001, Israel}
}
\date{\today}

\begin{abstract}
The Casimir force between macroscopic bodies is well understood, but not the Casimir force inside bodies. Guided by a physically intuitive picture, we develop the macroscopic theory of the renormalised Casimir stress inside planar materials (where the electromagnetic properties vary in one direction). 
\end{abstract}

\maketitle

\section{Introduction}

Casimir \cite{Casimir} and van der Waals \cite{vdW} forces are the subject of a mature research area \cite{Lifshitz,Schwinger,Milonni,Milton,BKMM,Leo,Review,Scheel} enjoying a renaissance \cite{Forces} since the first precision measurement of the Casimir force \cite{Lamoreaux}. Current theory \cite{Review} is able to predict the force between macroscopic bodies with an accuracy only limited by the knowledge of the material parameters. Yet surprisingly \cite{SimpsonSurprise} after almost 70 years of research, while theory describes the Casimir force \emph{between} bodies, the force \emph{inside} macroscopic bodies is poorly understood. In this paper, we develop from a physically intuitive picture the macroscopic theory of the stress that gives rise to Casimir forces inside inhomogeneous planar media. Our result agrees with a previous {\it ad hoc} procedure \cite{Philbin}, giving both mathematical and physical justification to a conjecture that was only tested numerically so far. Our theory is likely to be extendable; it may open the gate to the computation of Casimir forces inside arbitrary materials.

In piece--wise homogeneous media, the Casimir force acts only upon boundaries \cite{Piecewise}, {\it i.e.}\ between bodies. In inhomogeneous media, the Casimir effect may act inside the material. In solids, the Casimir stress is negligible in comparison with the interatomic forces, but in fluids it may build up sufficient pressure to force fluids to move until an equilibrium is reached. This internal Casimir force is particularly strong near edges in the refractive--index profile \cite{Paper2}. 

Our paper develops the theoretical tools for predicting how inhomogeneous fluids respond to Casimir forces. We consider the simplest non--trivial class of inhomogeneous materials: planar media. These are materials where the electric permittivity $\varepsilon$ and magnetic permeability $\mu$ vary in only one spatial direction. We also assume that both $\varepsilon$ and $\mu$ are scalars (isotropic media). Note that in using $\varepsilon$ and $\mu$ we are using the concepts of macroscopic electromagnetism \cite{LL8}; we do not need to resort to the microscopic properties of materials. Our theory thus indicates that macroscopic electromagnetism --- plus quantum fluctuations --- is sufficient to predict the Casimir force, which is far from being obvious.

\begin{figure}[t]
\begin{center}
\includegraphics[width=19.pc]{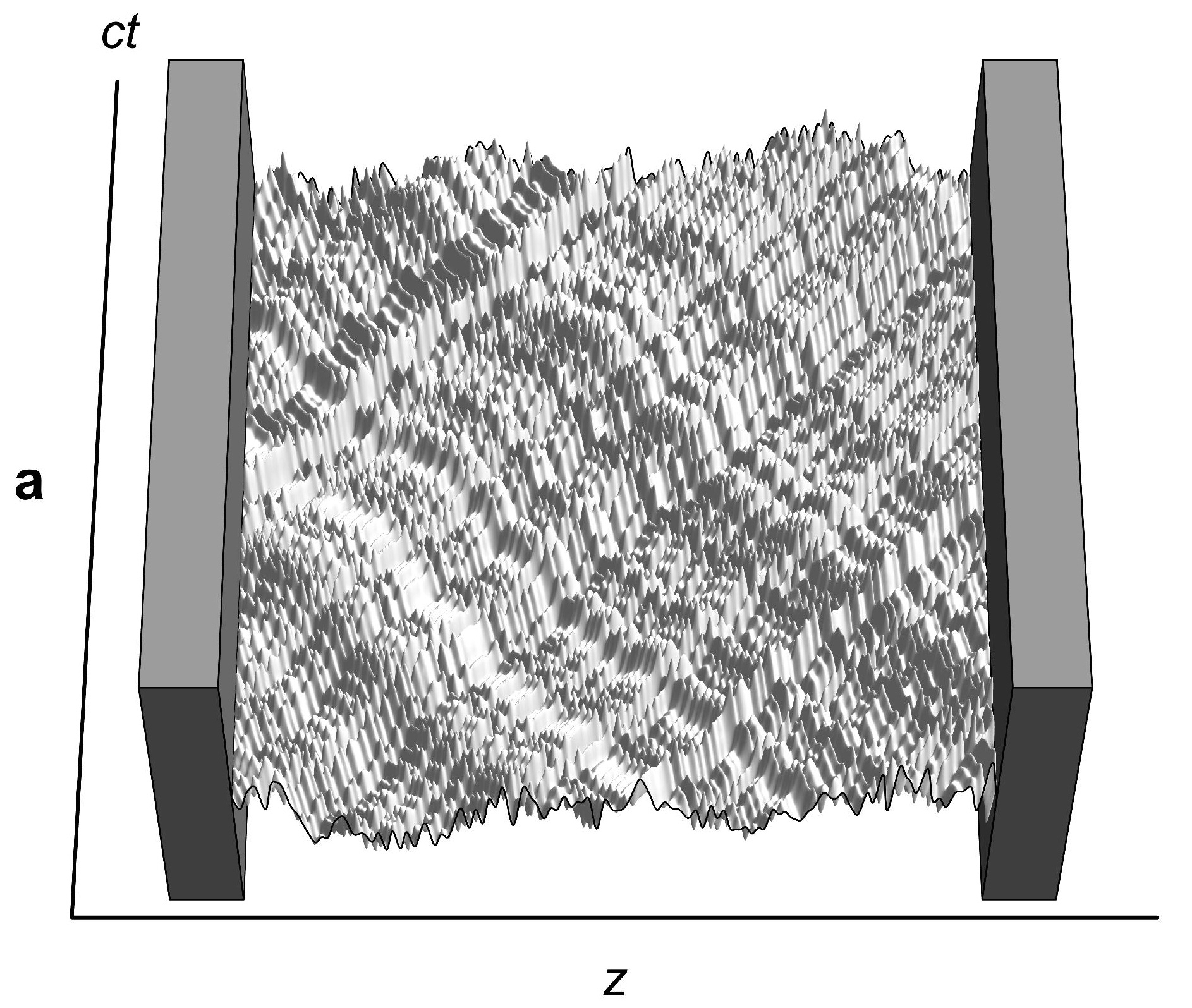}
\includegraphics[width=19.pc]{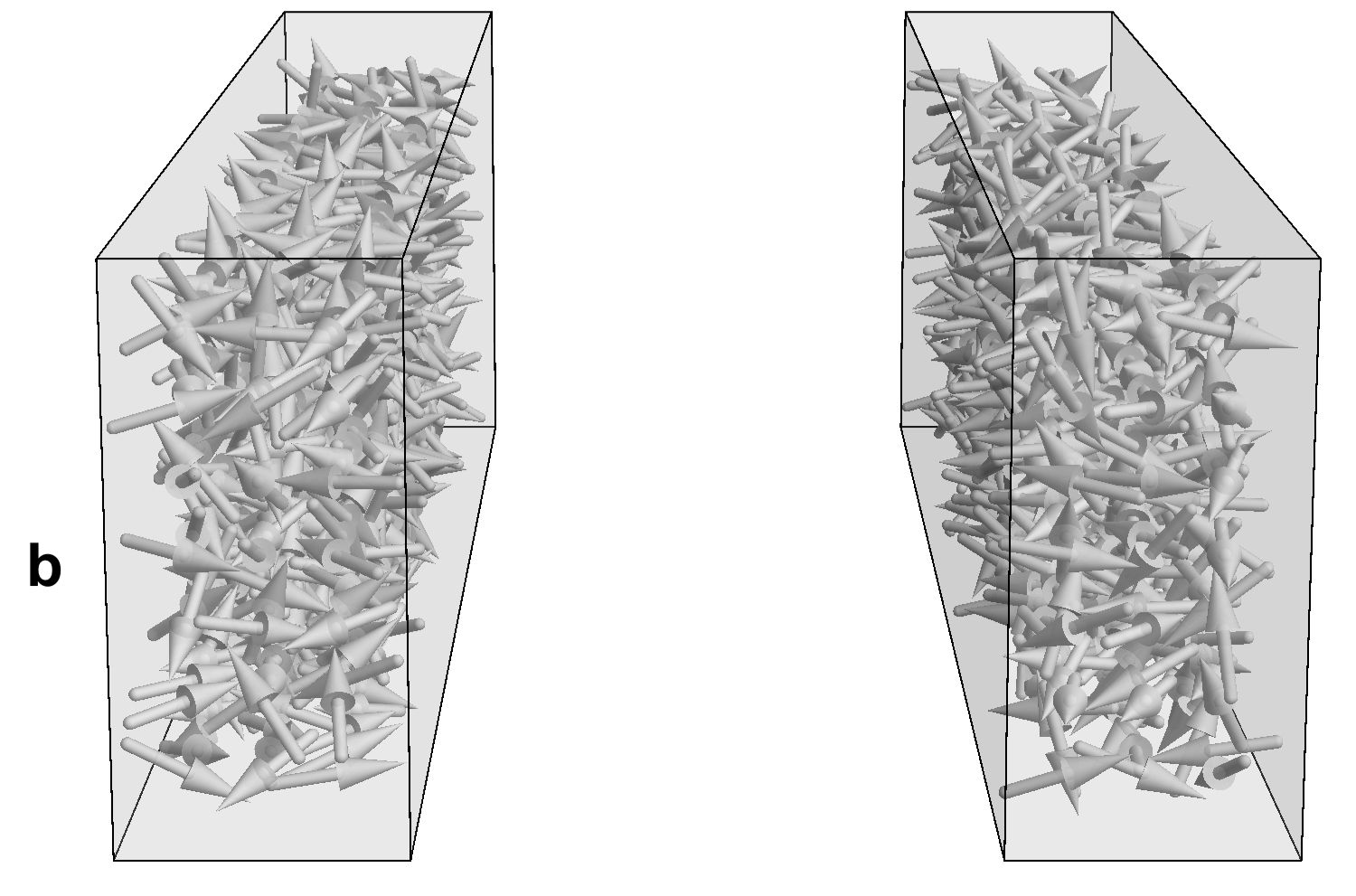}
\caption{
\small{
Visualization of quantum noise. {\bf a}: Casmir's picture: space--time diagram of the fluctuating field in the vacuum state. Boundaries create reflections, causing the zero-point energy to depend on them, which gives a force (as derivative of the energy with respect to the position of the boundary). {\bf b}: Schwinger's picture: vacuum fluctuations induce random dipoles in materials, causing them to interact with each other, which gives the same force.
}
\label{quantumnoise}}
\end{center}
\end{figure}

Let us begin by developing heuristic arguments and intuitive pictures for Casimir forces, visualizing the theoretical problems they pose and our solutions. The root of the problem has been the interplay between quantum fluctuations and electromagnetism that creates the Casimir force in the first place. Figure \ref{quantumnoise} illustrates two fundamentally different theoretical concepts of quantum fluctuations. One is Casimir's idea \cite{Casimir} --- inspired by Bohr \cite{Milonni,Scheel} --- that vacuum fluctuations of the electromagnetic field are responsible for the effect (Fig.~\ref{quantumnoise}a), the other Schwinger's concept \cite{Schwinger} --- Lifshitz' theory \cite{Lifshitz} is consistent with --- of fluctuating sources interacting with the electromagnetic field (Fig.~\ref{quantumnoise}b). In Casimir's picture (Fig.~\ref{quantumnoise}a) the zero--point energy of the electromagnetic quantum field in and around materials creates the force, in Schwinger's (Fig.~\ref{quantumnoise}b) the field is not even required to be quantized \cite{SchwingerSource}. Here, quantum fluctuations in the material induce oscillating dipoles at each point (Fig.~\ref{quantumnoise}b). Each dipole sends out an electromagnetic wave, the wave explores the surroundings, and when some part of the wave is scattered back to the source it interacts with it. In Schwinger's picture (Fig.~\ref{quantumnoise}b) the interaction of each fluctuating dipole with the backscattered wave creates the Casimir force.

Both pictures harbour similar hidden problems: infinities. In Casimir's picture (Fig.~\ref{quantumnoise}a) the zero--point energy $E$ sums up the vacuum energies of all the allowed electromagnetic modes with eigenfrequencies $\omega_m$:
\begin{equation}
E = \sum_m \frac{\hbar\omega_m}{2} = \infty \,.
\label{infinity}
\end{equation}
One sees this pictorially from trying to plot (Fig.~\ref{quantumnoise}a) the electric field strength in the vacuum state for Casimir's classic example \cite{Casimir}, the vacuum in the cavity between two perfectly reflecting mirrors \cite{Leo}. Adding more and more cavity modes produces finer and finer details in the fluctuating field, while the amplitude grows without bound. Yet Casimir \cite{Casimir} managed to extract the part from the diverging series in Eq.~(\ref{infinity}) that actually does physical work and found a finite force. Procedures that achieve this are called \emph{renormalisation}. In the zeta--function renormalisation \cite{BKMM,Zeta}, for example, the eigenfrequencies in Eq.~(\ref{infinity}) are raised to the power $1-s$ for complex $s$, which defines an analytic function of $s$ where the series converges. The analytic continuation to $s=0$ then assigns a finite value to Eq.~(\ref{infinity}) which happens to agree with Casimir's original calculation \cite{Casimir}.

\begin{figure}[h]
\begin{center}
\includegraphics[width=19.0pc]{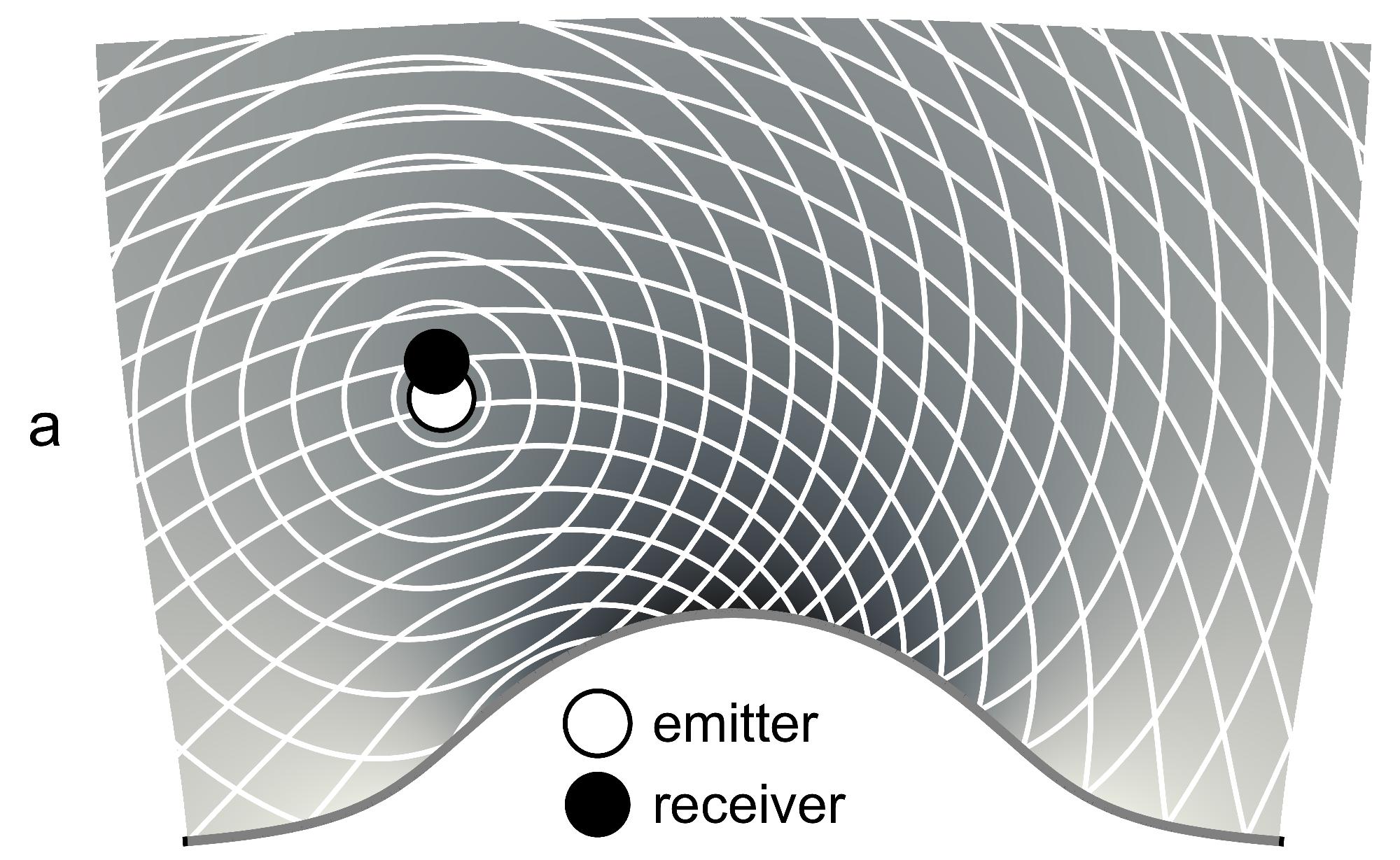}
\includegraphics[width=19.0pc]{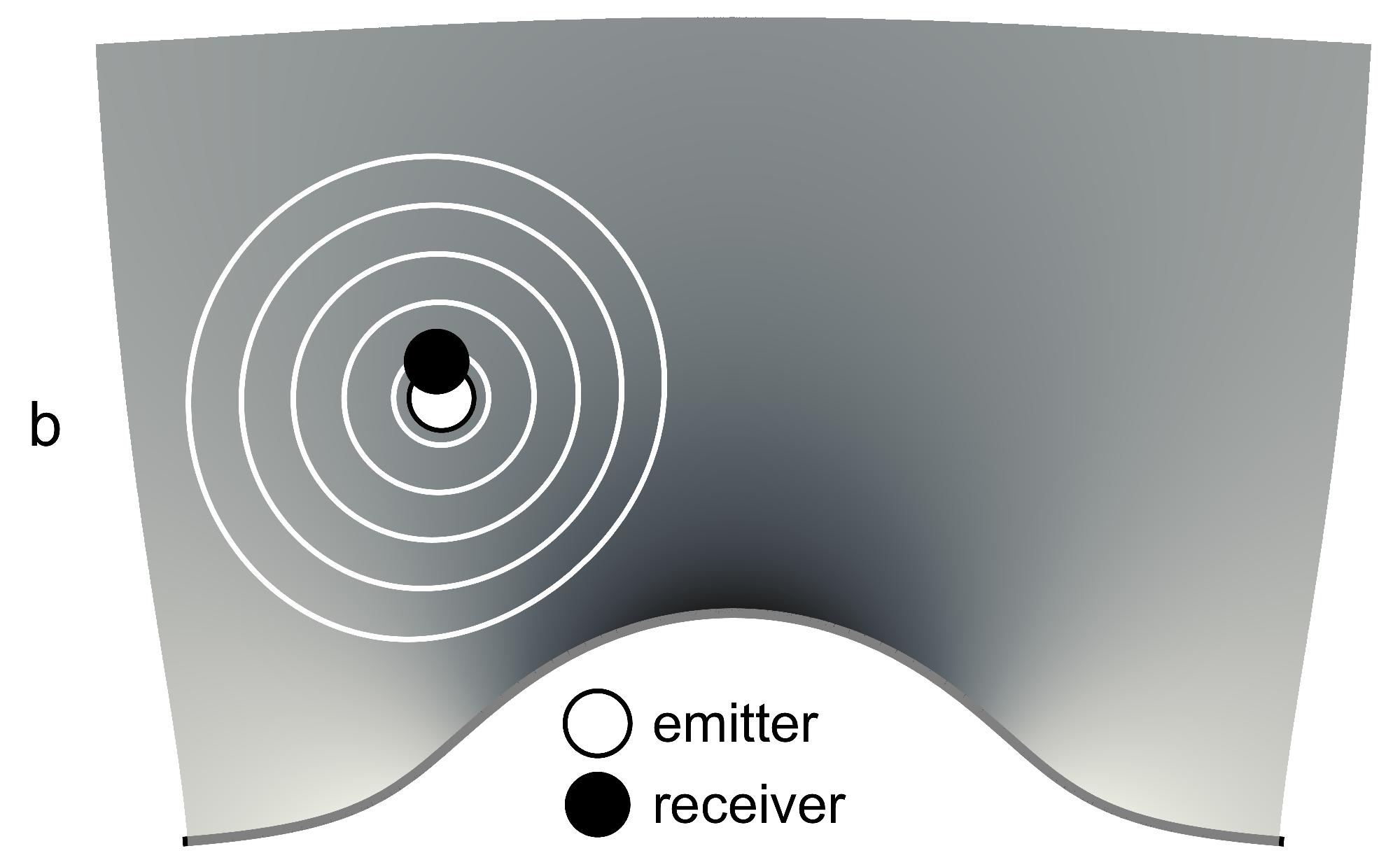}
\caption{
\small{
Point-splitting method. Each point of the inhomogeneous medium (different shades of grey) is split into emitter and receiver. {\bf a}: the emitter sends out waves with phase fronts indicted by white lines that the receiver picks up. The receiver should only respond to the scattered wave, but electromagnetic theory includes the direct interaction between emitter and receiver, illustrated in {\bf b}. In our renormalisation we subtract this outgoing wave. One sees that the outgoing wavefronts are deformed in the inhomogenous medium; the outgoing wave depends on the local environment, including its spatial variations. The picture shows a transformation medium \cite{GREEBook} with reflecting boundary (carpet cloak) where only the boundary scatters the light, whereas ordinary media scatter light inside.  
}
\label{pointsplit}}
\end{center}
\end{figure}

An infinity also mars Schwinger's picture of fluctuating sources (Fig.~\ref{quantumnoise}b), but we suggest one can identify and remove the issue in a physically motivated way (Fig.~\ref{pointsplit}). The infinity comes from the interaction of each dipole with itself, which poses already a problem in the classical theory of electromagnetism \cite{Jackson}. There one must explicitly forbid point charges to directly interact with themselves. In Lifshitz' \cite{Lifshitz} and Schwinger's \cite{Schwinger} theory each point of the medium is mentally split into two points, that we view as emitter and receiver (Fig.~\ref{pointsplit}a) infinitesimally close to each other. The quantum fluctuations of the emitter generate electromagnetic waves the receiver responds to. The contribution due to the wave directly going from emitter to receiver (Fig.~\ref{pointsplit}b) is deemed unphysical and removed. 

Lifshitz' \cite{Lifshitz} and Schwinger's \cite{Schwinger} theory works well for arbitrary piece--wise homogeneous planar materials \cite{Review}; it also agrees with Casimir's original result \cite{Leo} for the special case of a perfect cavity. However, the theory fails for inhomogeneous media \cite{Philbin,Simpson}, for the following reason. The outgoing wave does not only perceive the electromagnetic properties of the material at the point of emission, but also in its vicinity, even if emitter and receiver are infinitesimally close. The near field does not only depend on the local values of $\varepsilon$ and $\mu$, but also on their first spatial derivatives (see Sec.~II and Appendix A). This feature is ignored in Lifshitz' \cite{Lifshitz} and Schwinger's \cite{Schwinger} theory, although it appears to be inevitable in macroscopic electromagnetism \cite{LL8} where each mathematical point of the material is not a physical point, but rather represents the collective effect of a very great number of molecules generating the electromagnetic response of the material \cite{Fluids}.

\begin{figure}[h]
\begin{center}
\includegraphics[width=18.0pc]{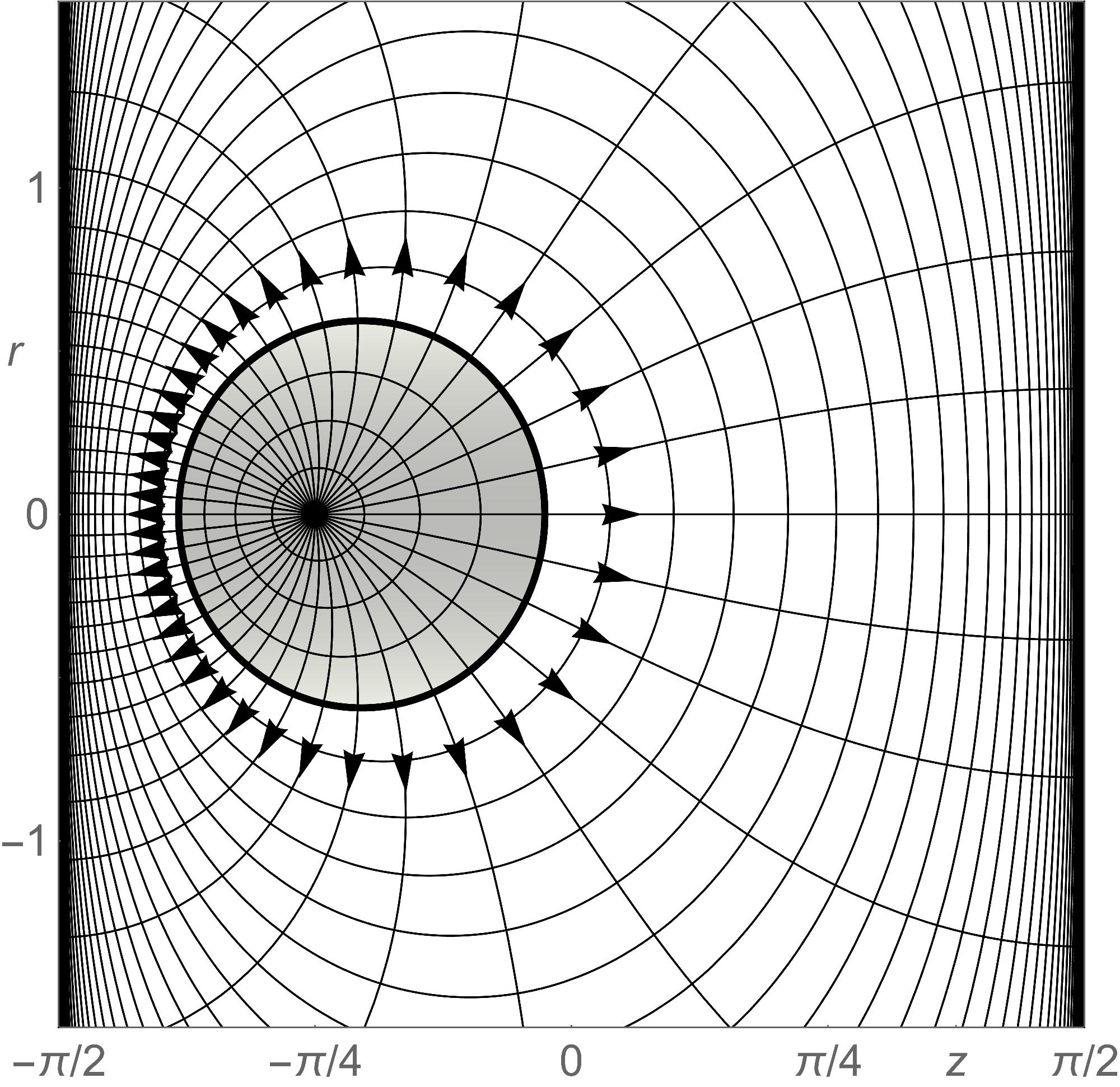}
\caption{
\small{
Outgoing wave. Rays (curves with arrows) and wave fronts (curves orthogonal to rays) outgoing from the emitter at various times. The figure also contains a snapshot at a fixed time that shows a negative image of the light flash (thick black curve) and of the amplitude (grey) lingering behind the flash due to geometric dispersion (Sec.~IIF). The figure employs the refractive--index profile (Fig.~\ref{quadratic}) of the example (Sec.~III) that depends on the $z$-coordinate; $r$ is the radius around the point of emission. The amplitude of geometric dispersion is given by Eq.~(\ref{bsec}).
}
\label{outgoing}}
\end{center}
\end{figure}

Now, if the dielectric properties of the material vary, both the outgoing wave and the scattered wave depends on them. How does one distinguish the outgoing wave? This is the theoretical problem in a nutshell. In this paper we apply geometrical optics and insights from transformation optics \cite{GREEBook} to identify the outgoing wave (Fig.~\ref{outgoing}). We achieve this for arbitrary planar $\varepsilon$ and $\mu$, not only for the case $\varepsilon=\mu$ when transformation optics is exact \cite{GREEBook}. This outgoing wave is then removed from the interaction between emitter and receiver. We prove (Appendix B) that this renormalization removes the principal divergences of the Casimir stress and we believe that the remaining stress is the complete physical one. 

\begin{figure}[h]
\begin{center}
\includegraphics[width=18.0pc]{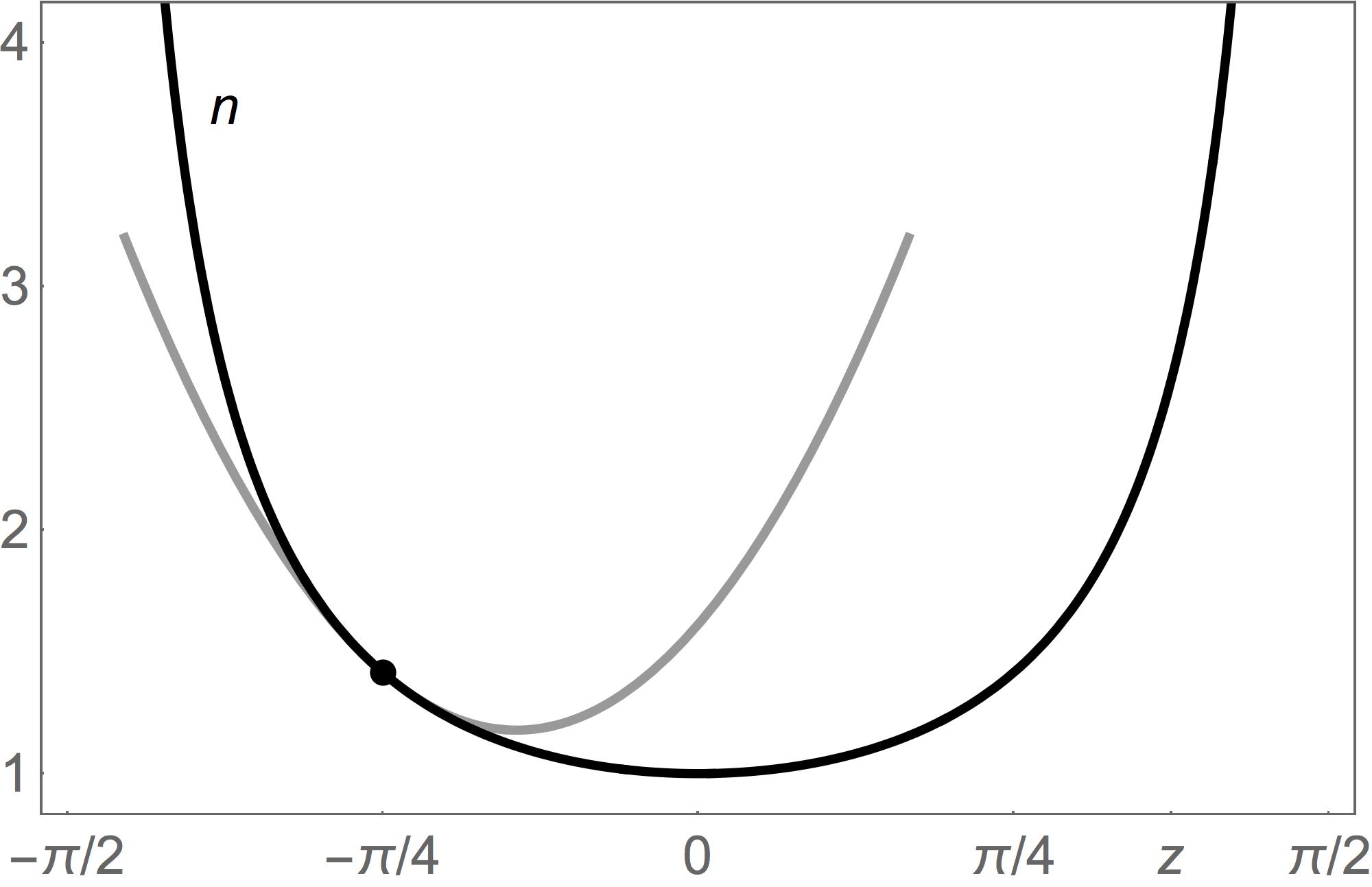}
\caption{
\small{
Quadratric expansion. The refractive--index profile (black curve) is quadratically expanded around the point of emission (grey parabola) in order to define the direct outgoing wave from emitter to receiver (Fig.~\ref{pointsplit}b). The figure illustrates the refractive index profile of the example, Eq.~(\ref{sec}) in Sec.~III.
}
\label{quadratic}}
\end{center}
\end{figure}

As the emitter perceives its environment beyond the immediate point of emission, the outgoing wave depends on the local dielectric properties and their derivatives. How many derivatives does one need to take into account? In this paper we use a quadratic expansion of the local material parameters (Fig.~\ref{quadratic}). We have seen that a linear expansion is insufficient (Appendix C); the quadratic expansion is the simplest one that works. Moreover, as waves are oscillations in space and time described by a second--order wave equation, one needs at least two infinitesimal steps to establish a wave, which might explain why one needs a quadratic expansion for identifying the outgoing wave.

Note that similar ideas have been proposed before in attempts \cite{Adler,Wald} to renormalize the energy--momentum tensor of the quantum vacuum in general relativity \cite{BD}. There the singularities of the emitted wave are removed by subtracting certain waves defined in quadratic expansion. This procedure did almost work, were it not for a logarithmic singularity. Now, one knows from transformation optics \cite{GREEBook} that certain magneto--electric materials with $\varepsilon=\mu$ correspond to space--time geometries in general relativity. So why does the same problem not occur in our theory? In classical general relativity \cite{LL2} the equivalence principle requires that the medium of space and time acts the same on all waves, regardless of wavelength: space--time is dispersionless. Yet ordinary materials are dispersive; $\varepsilon$ and $\mu$ depend on frequency $\omega$ and approach unity in the limit $\omega\rightarrow\infty$. We show in Appendix A that dispersion removes the logarithmic singularity in the stress. Moreover, in contrast to the previous papers \cite{Adler,Wald} we give simple explicit expressions for the regularizer that are ready for immediate use in numerical calculations. 

Dispersion has also been the problem in the {\it ad hoc} generalization \cite{Philbin} of Lifshitz theory \cite{Lifshitz} to inhomogeneous planar media \cite{Leo}. While our renormalizer turns out to be identical to the one proposed there \cite{Philbin}, the numerical examples \cite{Philbin} are incorrect, as they are for nondispersive media. In truth, the stress diverges there, but only logarithmically, which is not easy to see numerically (because ``the logarithm is a good approximation for a constant'' --- L. D. Landau). Moreover and more importantly, our paper gives mathematical proof beyond numerics that the renormalization works (see Appendix B) and adds a physical picture to the {\it ad hoc} renormalizer \cite{Philbin}, a picture that also seems valid beyond planar media.

Quantum field theory in curved space \cite{Adler,Wald,BD} and {\it ad hoc} renormalization \cite{Philbin} have not been the only attempts in developing a theory for the Casimir stress inside materials. Other theories have been proposed as well \cite{Goto,Bao,MiltonFulling,Churchill} including the perspective of going beyond standard macroscopic electromagnetism \cite{Churchill}. Zeta-function renormalization has been tried \cite{Bordag}, but as the authors put it \cite{Bordag}: ``The functional dependence of the needed counterterms is very involved and the classical model necessary for renormalization lacks physical intuition as well as motivation'', which is what we hope to facilitate in our paper. 

\section{Calculation}

In this section we substantiate the physical arguments and visualizations of the Introduction by theoretical calculations. We begin with a brief summary of Lifshitz theory in planar media.

\subsection{Lifshitz theory}

Consider a planar isotropic medium with scalar $\varepsilon$ and $\mu$ depending on the Cartesian coordinate $z$ and frequency $\omega$. We define the refractive index $n$ and impedance $Z$ as
\begin{equation}
n^2 = \varepsilon\mu \,,\quad Z^2 = \mu/\varepsilon \,.
\label{nZ}
\end{equation}
The Casimir--force density $\bm{f}$ inside the material is the divergence of the Casimir stress $\sigma$, the Minkowski stress in the vacuum state \cite{Pitaevskii}. For planar isotropic media the stress tensor is diagonal \cite{Leo} and hence only $\sigma_{zz}$ makes a contribution to the force: 
\begin{equation}
\bm{f} = \bm{\nabla} \cdot \sigma = (0,0,\partial_z\sigma_{zz})\,.
\label{force}
\end{equation}
According to Lifshitz theory \cite{Leo} the physically relevant stress component $\sigma_{zz}$ is given by the formula
\begin{equation}
\sigma_{zz} = \left. -\frac{\hbar c}{2\pi} \int_0^\infty \sum_{\mathrm{p}=\mathrm{E},\mathrm{M}} \frac{1}{\nu_\mathrm{p}}\left(w^2-\partial_z\partial_{z_0}\right) g_\mathrm{p}\,  \right|_{\bm{r}_0\rightarrow\bm{r}} 
\!\!\!\!\!\!\!\!\!\!\!d\kappa
\label{stress}
\end{equation}
where $c$ is the speed of light in vacuum, $\kappa$ denotes the imaginary wave number to the frequency $\omega = ic\kappa$, and $w^2$ abbreviates the expression 
\begin{equation}
w^2 = \varepsilon\mu\kappa^2 - \partial_x^2 - \partial_y^2 \,,
\end{equation}
while the $\nu_\mathrm{p}$ denote
\begin{equation}
\nu_\mathrm{E} = \mu \,,\quad \nu_\mathrm{M} = \varepsilon 
\label{nu}
\end{equation}
for the two polarizations E and M of the electromagnetic waves in the planar medium. The polarizations are defined as follows: in the E--polarization, the electric field vector lies in the plane orthogonal to $z$ (and is orthogonal to the wave vector), in the M--polarization it is the magnetic field. The scalar Green functions $g_\mathrm{p}(\bm{r},\bm{r}_0)$ satisfy the wave equation \cite{Leo}
\begin{equation}
\bm{\nabla}\cdot \frac{1}{\nu_\mathrm{p}} \bm{\nabla} g_\mathrm{p} - \frac{n^2\kappa^2}{\nu_\mathrm{p}} g_\mathrm{p} = \delta(\bm{r}-\bm{r}_0) \,.
\label{Green}
\end{equation}
They describe for the two polarizations the amplitudes of electromagnetic waves emitted at $\bm{r}_0$ and received at $\bm{r}$. According to the point--splitting method, one calculates the stress first for $\bm{r}_0\neq\bm{r}$ and then takes the limit $\bm{r}_0\rightarrow\bm{r}$. Note that the Green functions are real, because $\varepsilon$ and $\mu$ are real for purely imaginary frequencies \cite{Lifshitz}.

Close to the point of emission the field diverges. There the derivatives in Eq.~(\ref{Green}) outweigh $n^2\kappa^2$ such that we can ignore this term: electrostatics ($\kappa = 0$) dominates the near field. Close to the emission point we can also ignore any spatial variations of $\varepsilon$ and $\mu$; the Green equation (\ref{Green}) reduces to the Poisson equation with the solution
\begin{equation}
g_\mathrm{p} \sim -\frac{\nu_\mathrm{p}}{4\pi|\bm{r}-\bm{r}_0|}\quad\mbox{for}\quad \bm{r}\sim\bm{r}_0 \,,
\label{coulomb}
\end{equation}
which shows the principal, Coulomb--type singularity of the field at the point of emission. However, as the stress depends on derivatives of the Green functions, see Eq.~(\ref{stress}), further details of the near field become important. They call for more careful consideration, as follows. 

\subsection{Fourier method}

Let us Fourier--transform the Green functions with respect to the spatial coordinates $x$ and $y$ the electromagnetic functions $\varepsilon$ and $\mu$ do not depend on. We obtain from Eq.~(\ref{stress}) the stress:
\begin{eqnarray}
\sigma_{zz} &=& -\frac{\hbar c}{(2\pi)^2} \int_0^\infty \int_0^\infty {\cal W}\,u\,du\,d\kappa \,,
\nonumber\\
{\cal W} &=& \left. \sum_{\mathrm{p}=\mathrm{E},\mathrm{M}} \frac{1}{\nu_\mathrm{p}}\left(w^2-\partial_z\partial_{z_0}\right) \widetilde{g}_\mathrm{p}\, \right|_{z_0\rightarrow z} 
\label{fourier}
\end{eqnarray}
where ${\cal W}$ is the spectral stress density and $u$ denotes the amplitude of the $x$ and $y$ Fourier components (we have performed the integration over the angle), and
\begin{equation}
w^2 = \varepsilon\mu\kappa^2 +u^2 \,,
\label{w2}
\end{equation}
while the Fourier--transformed Green functions satisfy the differential equation
\begin{equation} 
\partial_z \frac{1}{\nu_\mathrm{p}} \partial_z \,\widetilde{g}_\mathrm{p} - \frac{u^2+n^2\kappa^2}{\nu_\mathrm{p}}\,\widetilde{g}_\mathrm{p} = \delta(z-z_0) \,.
\label{gdiff}
\end{equation}
Note that in calculating the stress with the Fourier method the limit $z_0\rightarrow z$ is taken first, before all integrations, as the notation of Eq.~(\ref{fourier}) indicates. 

The Lifshitz regularisation \cite{Lifshitz} assumes the medium to be homogenous in the vicinity of the emission point. For an infinitely extended homogenous medium, we obtain for Eq.~(\ref{gdiff}) the retarded solutions
\begin{equation}
\widetilde{g}_\mathrm{p} = - \frac{\nu_\mathrm{p}}{2w} \,\exp\left(-w|z-z_0|\right) \,.
\label{lifsh}
\end{equation}
Substituting the two $\widetilde{g}_\mathrm{p}$ into the expression for the stress, Eq.~(\ref{fourier}), we see that $\sigma_{zz}$ diverges like $\Lambda^4$ where $\Lambda$ is a cut--off of each of the integrals in Eq.~(\ref{fourier}). However, more careful analysis (Appendix A) reveals that the unrenormalized, bare stress diverges like
\begin{equation}
\sigma_{zz} \sim \alpha_4\Lambda^4 +  \alpha_2\Lambda^2 + \alpha \ln \Lambda
\label{singularities}
\end{equation}
with the terms
\begin{eqnarray}
\alpha_4\Lambda^4 &=& \frac{\hbar c}{(2\pi)^2} \int_0^\Lambda \int_0^\Lambda 2w\,u\,du\,d\kappa \,,
\label{a4}
\\
\alpha_2\Lambda^2 &=& -\frac{\hbar c}{(2\pi)^2} \int_0^\Lambda \int_0^\Lambda \frac{n^2 Z'^2w^4 + n'^2Z^2 u^4}{4n^2Z^2w^5}\, u\,du\,d\kappa \,.
\nonumber
\end{eqnarray}
The primes denote derivatives with respect to $z$. The coefficient $\alpha$ of the logarithm is complicated (Appendix A) but turns out to vanish for physical media with dispersion where $\varepsilon$ and $\mu$ depend on frequency and approach unity for $\kappa\rightarrow\infty$ (Appendix A). The other two singularities do not disappear in general, although $\alpha_2\Lambda^2$ becomes linear due to the $u$--integration, as the near field of the emitter is an evanescent field \cite{BornWolf} that contains arbitrarily large spatial Fourier components, even for finite frequencies. We recover the $\Lambda^4$ Lifshitz divergence, but we have also obtained a $\Lambda^2$ or $\Lambda$ singularity when $\varepsilon$ or $\mu$ depend on $z$. The latter is the mathematical reason why the original Lifshitz renormalization \cite{Lifshitz} fails in inhomogeneous planar media \cite{Philbin,Simpson}.

\subsection{\emph{Ad hoc} renormalisation}

In order to renormalize the Casimir stress one should subtract from the Green function $g_\mathrm{p}$ the near field $g_0$ of the emitter:
\begin{equation}
g \rightarrow g - g_0 \,.
\label{renorm}
\end{equation}
In the original Lifshitz renormalization \cite{Lifshitz} $\widetilde{g}_0$ is given by Eq.~(\ref{lifsh}). Philbin {\it et al.} \cite{Philbin} proposed to replace this by 
\begin{equation}
\widetilde{g}_\mathrm{0p} = - \frac{1}{2}\frac{\sqrt{\nu_\mathrm{p}(z)\nu_\mathrm{p}(z_0)}}{\sqrt{w_\mathrm{p}(z)w_\mathrm{p}(z_0)}} \,\exp\left(-\left|\int_{z_0}^z w\,dz\right|\right) \,.
\label{adhoc}
\end{equation}
Substituting $\widetilde{g}_{0p}$ into Eq.~(\ref{fourier}) for the stress one gets
\begin{equation}
\sigma_{zz} = \alpha_4\Lambda^4 +  \alpha_2\Lambda^2
\end{equation}
with exactly the terms of Eq.~(\ref{a4}) and hence all the singularites, apart from the logarithmic one. As the latter disappears for dispersive media, we have thus proved mathematically that this renormalizer does indeed remove all the physically relevant singularities of the Casimir stress in planar media.

One might be content with Eq.~(\ref{adhoc}) were it not for the unsettling thought that removing the infinity is necessary but not sufficient for identifying the correct renormalization. Subtracting one infinity from the other does not guarantee the right result, because the correct renormalizer might also make a finite contribution to the remaining finite stress, correcting it. In addition to the mathematical requirement of cancelling the singularities, a physical motivation is essential, which is what we are going to develop now. 

\subsection{Geometrical formulation}

Figure \ref{pointsplit} shows the physical picture we have in mind: the renormalizer removes the wave directly going from emitter to receiver, because this wave describes the unphysical interaction of the dipole with itself (mentally split into emitter and receiver). Given that the near field depends on both the values and the derivatives of the material parameters, as Eqs.~(\ref{singularities}-\ref{a4}) indicate, how can we distinguish the outgoing from the scattered wave? 

Scattering is a deviation from geometrical optics, because in geometrical optics a flash of light would just propagate along the ray trajectories; nothing is reflected there --- nothing is scattered.  Hence we identify the outgoing wave by geometrical optics. Since the defining feature --- the flash of light --- is a feature of wave propagation in physical, three-dimensional space, we consider geometrical optics in real space, not in Fourier space, but then transform our result for use in the Fourier method (Appendix B). 

First we cast the wave equation in geometrical form. In the rest of the paper we drop the polarization index p for keeping the notation uncluttered. We represent the scalar Green functions as
\begin{equation}
g = \nu(z)\,\nu(z_0)\,D
\label{ddef}
\end{equation}
and obtain from Eq.~(\ref{Green}) the wave equation:
\begin{equation}
\frac{\bm{\nabla}\cdot\nu\bm{\nabla} D}{n^2\nu} - \frac{R_z^z}{2}\,D - \kappa^2 D = \frac{\delta(\bm{r}-\bm{r}_0)}{n^2\nu} 
\label{dwave}
\end{equation}
with the abbreviation
\begin{equation}
R_z^z = \frac{2\left(\bm{\nabla}\nu\right)^2}{n^2\nu^2} - \frac{2\bm{\nabla}^2\nu}{n^2\nu} \,.
\label{ricci}
\end{equation}
For $\varepsilon=\mu=n$ the medium establishes an exact spatial geometry \cite{GREEBook} with the infinitesimal optical length $ds$ as line element, 
\begin{equation}
ds^2 = n^2\left(dx^2+dy^2+dz^2\right) \,.
\label{ds}
\end{equation}
In this geometry of light the term $n^{-3}(\bm{\nabla}\cdot n \bm{\nabla} D)$ in the wave equation (\ref{dwave}) is the Laplacian of the scalar $D$, while $R_z^z$ is the $zz$--component of the Ricci tensor \cite{GREEBook}. Equation (\ref{dwave}) describes the propagation of one polarization component taken from the conformally coupled wave equation of the electric or magnetic field \cite{GREEBook}.

\subsection{Geometrical optics}

After having given the wave equation a geometrical appearance, it is elementary to find the form of the Green function in geometrical optics. Let us write $D$ in terms of the amplitude ${\cal A}$ and phase $\varphi$ where we, for reasons of convenience, consider real wavenumbers $k$ and make the transition to imaginary $k=i\kappa$ later. We thus put
\begin{equation}
D = {\cal A} \,e^{i\varphi}
\end{equation}
and obtain from the real part of Eq.~(\ref{dwave}):
\begin{eqnarray}
\frac{\bm{\nabla}\cdot\nu\bm{\nabla} {\cal A}}{n^2\nu} &=& \Xi \,{\cal A} +  \frac{\delta(\bm{r}-\bm{r}_0)}{n^2\nu} \,,
\label{amplitude}\\
\Xi &=& \frac{\left(\bm{\nabla\varphi}\right)^2}{n^2} - k^2 + \frac{R_z^z}{2} \,,
\label{eikonal0}
\end{eqnarray}
while we get for the imaginary part:
\begin{equation}
\bm{\nabla}\cdot\left(\nu{\cal A}^2\bm{\nabla}\varphi\right) = 0\,.
\label{cont}
\end{equation}
The latter, an equation of continuity, describes the transport of photons from the point of emission along the gradient of the phase.

In geometrical optics, ``phase rules the waves''; the dominant feature of wave propagation is phase propagation. In a formal sense, geometrical optics gives the asymptotics $D_0$ for $D$ in the limit of large wavenumbers $k$ as
\begin{equation}
D_0 = {\cal A}_0 \,e^{iks}
\label{d0}
\end{equation}
with ${\cal A}_0$ being independent of $k$. In leading order, we obtain from Eqs.~(\ref{amplitude}-\ref{eikonal0}) the eikonal equation
\begin{equation}
\left(\bm{\nabla}s\right)^2 = n^2
\label{eikonal}
\end{equation}
showing that $s$ is the optical path length as perceived by the geometry of light, Eq.~(\ref{ds}). The amplitude ${\cal A}_0$ is then entirely enslaved to the phase as the solution of the equation of continuity, 
\begin{equation}
\bm{\nabla}\cdot\left(\nu{\cal A}_0^2\bm{\nabla}s\right) = 0
\label{cont0}
\end{equation}
with, from Eqs.~(\ref{coulomb}) and (\ref{ddef}), the condition
\begin{equation} 
{\cal A}_0 \sim -\frac{1}{4\pi\nu\,|\bm{r}-\bm{r}_0|}\quad\mbox{for}\quad \bm{r}\sim\bm{r}_0 \,.
\label{a0}
\end{equation}
Although the geometrical $D_0$ is introduced in a formal sense as an approximation, $D_0$ turns out to describe correctly the divergence of the full $D$ wave in the vicinity of the emission point (Appendix B). Additionally, we see from Fourier--transforming $D_0$ of Eq.~(\ref{d0}) with respect to time that $D_0$ corresponds to the flash of light ${\cal A}_0\,\delta(ct-s)$ along the geodesics, which shows that $D_0$ is the outgoing wave. 

\subsection{Geometric dispersion}

Geometrical optics gives the outgoing wave, but it turns out we also need to consider the next--order correction. For finding it, we write the phase as 
\begin{equation} 
\varphi \sim ks + \beta/k + ...
\label{phi}
\end{equation}
and consider the leading correction $D_1$ to the $D_0$ of geometrical optics:
\begin{equation} 
D \sim D_0 + D_1\,,\quad D_1 = -e^{iks}\frac{b}{ik} \,,\quad b = {\cal A}_0 \beta \,.
\label{b}
\end{equation}
To first order in $k^{-1}$ we obtain from Eqs.~(\ref{amplitude}), (\ref{eikonal0}), (\ref{eikonal}) and (\ref{phi}) the expression:
\begin{eqnarray} 
2\beta_1 &=& \frac{\bm{\nabla}\cdot\nu\bm{\nabla} {\cal A}_0}{n^2\nu {\cal A}_0} - \frac{R_z^z}{2} \quad\mbox{for}\quad  \bm{r}\neq\bm{r}_0 \quad\mbox{with}
\nonumber\\
\beta_1&=& \frac{\bm{\nabla}s\cdot\bm{\nabla}\beta}{n^2} \,.
\label{beta1}
\end{eqnarray}
Assuming $\beta$ is a function of $s$ we see from Eq.~(\ref{eikonal}) that $\beta_1$ is the derivative of $\beta$ with respect to $s$, so $\beta$ is the integral of $\beta_1$. As $\beta_1$ is finite for $\bm{r}\rightarrow\bm{r}_0$ the integral is well-behaved at the point of emission where the phase and hence $\beta$ vanishes. Consequently, we can write
\begin{equation} 
\beta = \int_0^s \beta_1 \,ds \,.
\label{beta}
\end{equation}
From this also follows that $b={\cal A}_0\beta$ is finite for $\bm{r}\rightarrow\bm{r}_0$ despite the amplitude ${\cal A}_0$ diverging there. 

How does the leading correction to geometrical optics, $D_1$, behave in time? The factor $1/(ik)$ in Fourier space corresponds to an integration in time of the delta function we obtained in Sec.~IIE, which gives a step function $\Theta$. Causality requires that the wave vanishes outside the light cone, so the Fourier--transformed $D_1$ is $\Theta(ct-z)\,b$. Due to geometric dispersion, the wave lingers with amplitude $b$ after the outgoing flash has passed. Of course, higher--order corrections will then let the wave recede. 

Note that geometric dispersion is the primary case of scattering in an inhomogeneous medium without sharp boundaries, as scattering is the deviation of wave propagation from geometrical optics and $D_1$ is the leading correction; $D_1$ describes the first reflection the outgoing wave encounters. According to Huygen's Principle \cite{BornWolf}, we can view the wave propagation as a continuous succession of propagations ($D_0$) and reflections ($D_1$). When geometrical dispersion vanishes so does scattering altogether.

\subsection{Quadratic expansion}

In the renormalisation of the Casimir force, we subtract the direct wave between emitter and receiver (Fig.~\ref{pointsplit}). This wave is sensitive to the local environment around the point of emission, as the singularities of the stress show, see Eq.~(\ref{a4}). They depend on $\varepsilon$ and $\mu$, and their first spatial derivatives. Therefore one might be tempted to expand $\varepsilon$ and $\mu$ to linear order, fit the outgoing wave to this expansion, calculate the resulting stress and subtract it from the bare stress, in the hope of getting a finite result. We tried it, it did not work (Appendix C). A quadratic expansion is needed, which also fits quite naturally the physical picture of wave propagation: waves are oscillations in space and time; to establish a wave one needs two spatial increments, hence quadratic expansion. We thus expand $n$ as
\begin{equation} 
n = n_0 + n_0'(z-z_0) + \frac{n_0''}{2}(z-z_0)^2
\label{quadex}
\end{equation}
and similarly $\nu$. In the following we use cylindrical coordinates $\{r,\phi,z\}$ where the point of emission lies on the $z$--axis ($r=0$), for taking advantage of the cylindrical symmetry of the outgoing wave. In a homogeneous medium, the wave would also be spherically symmetric with optical path length $s = n_0 \rho$ where $\rho$ denotes the spherical radius,
\begin{equation} 
\rho = \sqrt{r^2+(z-z_0)^2} \,.
\end{equation}
In an inhomogeneous medium, we determine the leading corrections to $s$, assuming $s$ to be proportional to $\rho$ with quadratic expansion of the proportionality factor in $r$ and $z-z_0$. For $r=0$ the geodesic is a straight line in $z$--direction, and so the geodesic length is simply the $z$--integral of $n$ in quadratic order, Eq.~(\ref{quadex}) integrated. For $r\neq 0$ we need to solve the eikonal equation, Eq.~(\ref{eikonal}), to quadratic order, and get
\begin{equation} 
s = \rho \left(n_0 + \frac{n_0'}{2}(z-z_0) + \frac{n_0''}{6} (z-z_0)^2 - \frac{n_0'^2 r^2}{24n_0}\right) \,.
\label{squad}
\end{equation}
The $r$--dependent term in Eq.~(\ref{squad}) describes the correction to the geodesic length $s$ for small $r>0$.

The amplitude ${\cal A}_0$ we find in a similar way: knowing that in homogeneous media ${\cal A}_0$ is given by $-(4\pi\nu\rho)^{-1}$ we postulate quadratic corrections to this principal dependence. We obtain from Eq.~(\ref{cont0}) and our previous result, Eq.~(\ref{squad}), to quadratic order:
\begin{equation} 
{\cal A}_0 = -\frac{1}{4\pi\sqrt{\nu_0\nu}}\left(\frac{1}{\rho} + \frac{n_0^2 R}{48}\rho\right)
\label{ampex}
\end{equation}
where $R$ is the curvature scalar \cite{GREEBook} in 3D:
\begin{equation} 
R = -\frac{4n_0''}{n_0^3} + \frac{2n_0'^2}{n_0^4} \,.
\label{curvature}
\end{equation}
We see that, to quadratic order, the outgoing wave in geometrical optics depends entirely on geometric quantities --- the geodesic length and the curvature scalar, apart from a mere prefactor containing $\varepsilon$ for the $M$ and $\mu$ for the $E$--polarisation. The stress of the wave is calculated in Appendix B. 

Now we turn to the amplitude $b$ of geometric dispersion. Using a quadratic expansion of $\varepsilon$ and $\mu$ we can only deduce $b$ to zeroth order, as Eq.~(\ref{beta1}) depends on second derivatives. From our results, Eqs.~(\ref{squad}-\ref{curvature}), follows
\begin{equation} 
\beta_1 = \frac{1}{24n_0^2}\left(\frac{n_0'^2-2n_0n_0''}{n_0^2} + \frac{6\nu_0\nu_0''-9\nu_0'^2}{\nu_0^2}\right)
\label{betaresult}
\end{equation}
for $z\rightarrow z_0$ and $r\rightarrow 0$ independent of the order of limits. We then obtain from Eqs.~(\ref{beta}) and (\ref{squad}-\ref{ampex}) to zeroth order:
\begin{equation} 
b = - \frac{n_0}{4\pi\nu_0}\,\beta_1 \,.
\label{bresult}
\end{equation}

For media with $\varepsilon=\mu=n$, establishing exact geometries \cite{GREEBook}, we see that $b$ is proportional to $n_0n_0''-2n_0'^2$. When does geometric dispersion cease to exist, when is geometrical optics exact? The general solution of the differential equation $n''n=2n'^2$ is $n=a(z-z_1)^{-1}$ with some constants $a$ and $z_1$ of the dimension of a length. A geometry with this $n$ in the line element of Eq.~(\ref{ds}) is the Beltrami space in 3D \cite{Needham}, the only maximally symmetric curved space \cite{Maximal} for planar $n$. Consequently, for planar media with $\varepsilon=\mu$, scattering vanishes if and only if $n$ establishes a maximally symmetric space.

There is another interesting aspect of geometric dispersion, one that is immediately relevant to the Casimir effect. Substitute in the Casimir stress, Eq.~(\ref{stress}), $g_\mathrm{p}=\nu_0^2D_1$ with $D_1=e^{-\kappa s} b/\kappa$ and $b$ given by Eq.~(\ref{bresult}). One finds that the stress vanishes exactly, if we take the limit  $z_0\rightarrow z$ first and $r\rightarrow 0$ later, as we should do in the Fourier method. Geometric dispersion causes no Casimir force. Furthermore, as geometric dispersion describes the first instance of scattering the outgoing wave encounters --- the first reflection --- we realise that the Casimir force does not depend on primary reflections in the material, only on secondary ones, and on multiple reflections. Recall Lifshitz' famous formula \cite{Lifshitz} for the Casimir stress between planar mirrors \cite{Leo}: it only depends on the product of the reflection coefficients of the two mirrors, not on single reflections.  Our result generalises this feature to arbitrary planar media. 

Note however, that for a different order of limits, for example if we take $r\rightarrow 0$ first and then consider $z\sim z_0$ later, $e^{-\kappa s} b/\kappa$ does make a contribution to the stress, a contribution that diverges for $z_0\rightarrow z$ (but not for $\kappa\rightarrow 0$). This contribution needs to be taken into account if the Casimir stress is renormalised in other procedures than  the Fourier method, for example by subtracting $g_0$ in physical space. 

\section{Example}

In this section we study an example (Fig.~\ref{quadratic}) that allows us to derive analytic expressions for the central ingredients of our theory, the outgoing wave in geometrical optics and the geometric dispersion (Fig.~\ref{outgoing}). We are not concerned whether this example is practically relevant --- examples elsewhere \cite{Paper2} are --- it should only allow us to derive analytic expressions where we can compare the approximations of Sec.~IIG with exact results. The example is geometrical in nature: it is the planar refractive--index profile that implements constant negative curvature (in a sense made precise below). 

\subsection{Constant curvature}

We assume that the medium establishes an exact spatial geometry for electromagnetic waves, {\it i.e.} we require \cite{GREEBook}:
\begin{equation} 
\epsilon = \mu = n(z) \,.
\label{exactgeo}
\end{equation}
Additionally, we require the $R_z^z$ component of the Ricci tensor \cite{GREEBook}, Eq.~(\ref{ricci}), to be constant:
\begin{equation} 
R_z^z = -\frac{2n''}{n^3} + \frac{2n'^2}{n^4} = \mbox{const} \,.
\label{rzz}
\end{equation}
Consider a radial plane around the $z$--axis, {\it i.e.} the plane $\{r,z\}$ with $\phi=\mbox{const}$ in cylindrical coordinates. 
On radial planes, the curvature scalar in 2D agrees with the Ricci component $R_z^z$ in 3D for $n=n(z)$. In 2D we can use visual intuition for curved spaces and tools from conformal mapping \cite{Needham}. We know, for instance, that the surface of the sphere is a space of constant 2D curvature. We also know that we can represent the sphere in stereographic projection \cite{GREEBook} by the refractive--index profile of Maxwell's fish eye \cite{GREEBook,Maxwell}. With conformal mapping, using the exponential map \cite{GREEBook}, we can map Maxwell's fish eye to the refractive--index profile of Mikaelian's lens \cite{Mikaelian}:
\begin{equation} 
n = \mathrm{sech} z 
\label{sech}
\end{equation}
where we rescaled $z$ such that it is dimensionless. One easily obtains from Eq.~(\ref{rzz}) that $R_z^z=+2$, so the hyperbolic--secant profile of the Mikaelian lens implements a space of positive constant curvature. Now, if we replace $z$ by $iz$ we get a negative curvature in Eq.~(\ref{rzz}), provided of course that $n$ remains real, which it does for $n=\mathrm{sech}(iz)=\mathrm{sec}z$. We obtain $R_z^z=-2$ for 
\begin{equation} 
n = \mathrm{sec} z \,.
\label{sec}
\end{equation}
In the hyperbolic secant of Eq.~(\ref{sech}) we can also shift $z$ by $i\pi/2$ and get $-i\mathrm{csch}z$. The imaginary prefactor changes the sign in Eq.~(\ref{rzz}): 
\begin{equation} 
n = \mathrm{csch} z 
\label{csch}
\end{equation}
defines a second profile of constant negative curvature with $R_z^z=-2$. The secant profile (Fig.~\ref{quadratic}) is confined between $\{-\pi/2,\pi/2\}$ where $\mathrm{sec} z$ diverges, but it represents an open space, because light will never reach the rim at $\{-\pi/2,\pi/2\}$, as it becomes infinitely slow. The hyperbolic-cosecant profile also establishes an open space, although being confined to $\{0,\infty\}$, because it diverges for $z\rightarrow 0$. In contrast, the hyperbolic secant corresponds to a closed space --- the surface of the sphere. There, for $z\rightarrow\pm\infty$, light becomes infinitely fast, reaching in a finite time infinity (and beyond).

One can multiply these three profiles with constant factors, rescale and shift $z$, which give three--parameter solutions to Eq.~(\ref{rzz}), a second--order differential equation with one parameter, the constant on the right--hand side. We have obtained the complete solution. This solution includes, as limiting cases, the profiles $n_1\exp(-z/a)$ of zero curvature, {\it i.e.} coordinate--transformed flat space \cite{GREEBook}, and also the Beltrami plane \cite{Needham}:
\begin{equation} 
n = \frac{1}{z} \,.
\label{beltrami}
\end{equation}
The latter is not only a space of constant curvature in 2D, but a maximally symmetric space \cite{Maximal} in 3D. In the following we focus on the secant profile of Eq.~(\ref{sec}), our test case.

\subsection{Green function}

First we calculate the Fourier--transformed Green function as the causal solution of the inhomogeneous differential equation, Eq.~(\ref{gdiff}), with $\epsilon = \mu = \nu = n(z)$ and $n(z)$ given by Eq.~(\ref{sec}). We apply the Wronskian method (Appendix A) where $\widetilde{g}$ is the product of two solutions of the homogeneous equation divided by their Wronskian. The two solutions are interchanged at $z=z_0$, which produces the jump in the derivative of $\widetilde{g}$ required for generating the delta function on the right--hand side of Eq.~(\ref{gdiff}). The homogeneous solutions are to be chosen such that one decays for $z\rightarrow-\pi/2$ and the other for $z\rightarrow+\pi/2$. They are the Legendre functions \cite{Erdelyi} $\mathrm{P}_\nu^{-\kappa}(\pm \sin z)$ with index
\begin{equation} 
\nu = \frac{1}{2} \left(-1 + \sqrt{1-4u^2}\right)
\label{nudef}
\end{equation}
not to be confused with the other $\nu$ defined in Eq.~(\ref{nu}). The Legendre functions reflect the pedigree of the secant profile that originates from the imaginary sphere. The waves on the sphere are the spherical harmonics \cite{Erdelyi} that are made of Legendre functions. Note we use the Legendre functions on the branch cut; they are defined by Eq.~3.4(6) of Ref.~\cite{Erdelyi}. We calculate the Wronskian using the asymptotics of the Legendre functions \cite{Erdelyi} and find for the Green function:
\begin{equation} 
\widetilde{g} = K_\nu^\kappa\,\mathrm{P}_\nu^{-\kappa}(\sin z)\,\mathrm{P}_\nu^{-\kappa}(-\sin z_0)
\label{gg}
\end{equation}
for $z>z_0$ (otherwise $z$ and $z_0$ are interchanged) with prefactor 
\begin{equation} 
K_\nu^\kappa = -\frac{\Gamma(\kappa-\nu)\Gamma(\kappa+\nu+1)}{4\pi}
\label{ggk}
\end{equation}
in terms of the Gamma functions \cite{Erdelyi}. Note that $\widetilde{g}$ is real, as it must by definition, Eq.~(\ref{gdiff}), even if the index $\nu$ of Eq.~(\ref{nudef}) is complex (for $u>1/2$). In this case $\kappa + \nu +1$ is the complex conjugate of $\kappa -\nu$, and so the prefactor $K_\nu^\kappa$ is proportional to $|\Gamma(\kappa-\nu)|^2$, which is real. One obtains from Eqs.~2.8(1) and 2.8(2) of Ref.~\cite{Erdelyi} that also $\mathrm{P}_\nu^{-\kappa}$ is real. 

\begin{figure}[h]
\begin{center}
\includegraphics[width=19.0pc]{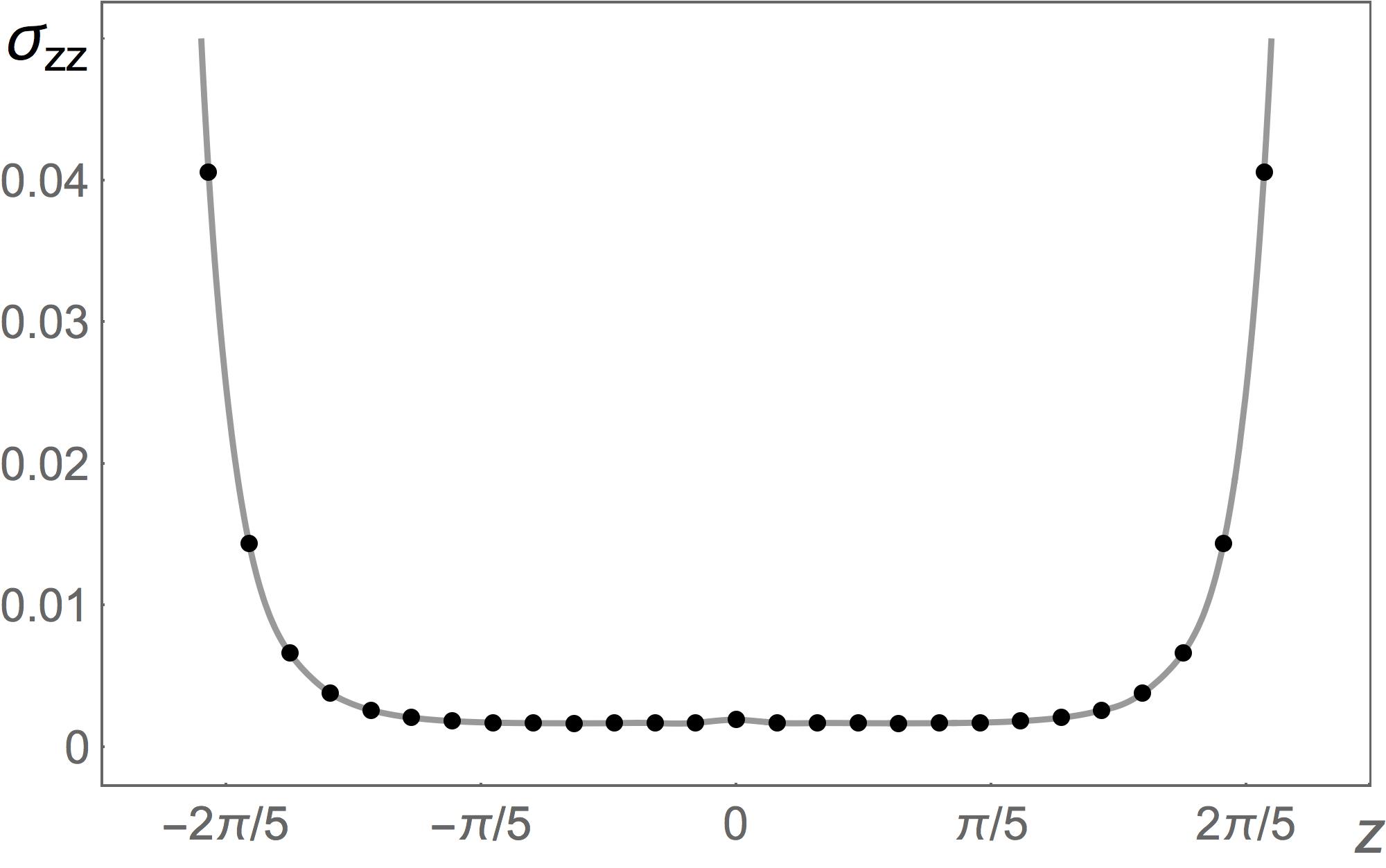}
\caption{
\small{
Numerical example. Casimir stress $\sigma_{zz}$ in units of $\hbar c$ in the secant profile (Fig.~\ref{quadratic}) as a function of $z$. The stress generates a repulsive force, Eq.~(\ref{force}); the stress rises significantly where $n(z)$ grows. Dispersion was included in the model by using $n=\mathrm{sec}(z/a)$ with $a=\sqrt{1+(\kappa/\kappa_0)^2}$ and $\kappa_0=3.0$. The stress was computed numerically (dots) using the Green function of Eqs.~(\ref{nudef}-\ref{ggk}) and our renormalisation procedure, Eq.~(\ref{renormalisation}), requiring subtractions with 50-digit precision, 30-digit accuracy and 500-digit maximal extra precision.}
\label{itay}}
\end{center}
\end{figure}

We found in Sec~IIG that for planar media establishing exact geometries, Eq.~(\ref{exactgeo}), only the profiles of maximally symmetric spaces are scatteringless. These are Beltrami spaces with the $1/z$ profile of Eq.~(\ref{beltrami}) scaled and shifted, not our case of the secant profile, Eq.~(\ref{sec}). Therefore our profile scatters electromagnetic waves \cite{LLRemark}; it generates Casimir stress (Fig.~\ref{itay}).

\subsection{Geodesic length}

The renormaliser of the stress is made by the outgoing wave in geometrical optics that depends on the geodesic distance $s$ from the point of emission. For the secant profile we can calculate $s$ analytically for arbitrary distances, not only in the infinitesimal environment of the emitter, as follows. First we note that $s$ does not depend on the angle $\phi$ due to cylindrical symmetry (the emitter sits at $r=0)$. The geodesics lie on the radial $\{r,z\}$ planes with metric
\begin{equation} 
ds^2 = n^2(dr^2 + dz^2) \,.
\label{ds2d}
\end{equation}
One can find the geodesics using conformal mapping with the exponential map and M\"{o}bius transformations \cite{GREEBook,Needham}; the details are not important here --- one verifies by elementary calculation that the coordinates $\xi$ and $\eta$ defined as
\begin{eqnarray} 
\xi &=& \frac{\sin z - \cosh r \, \sin z_0}{\cosh r + \cos(z+z_0)}  \,,
\nonumber\\
\eta &=& \frac{\sinh r \, \cos z}{\cosh r + \cos(z+z_0)} 
\label{xieta}
\end{eqnarray}
express the line element of Eq.~(\ref{ds2d}) as $ds^2=n_p^2(d\xi^2+d\eta^2)$ where
\begin{equation} 
n_p = \frac{2}{1-\xi^2-\eta^2} \,.
\label{poincare}
\end{equation}
This is the refractive--index profile of the Poincar\'{e} disk \cite{Needham}. We obtain from Eq.~(\ref{xieta}) for the radius $\sigma$ on the disk
\begin{equation} 
\sigma = \sqrt{\xi^2+\eta^2} = \frac{c_-}{c_+}
\label{sigma}
\end{equation}
with the abbreviations
\begin{equation} 
c_\pm = \sqrt{\cosh r \pm \cos(z\pm z_0)} \,.
\label{cpm}
\end{equation}
We see that $\sigma$ vanishes at the point of emission where $r=0$ and $z=z_0$. With the transformation of Eq.~(\ref{xieta}) we have thus managed to put the emitter in the center of the Poincar\'{e} disk. The geodesics from the center are simply the radial lines. Hence we obtain for the geodesic length:
\begin{equation} 
s = \int n_p \,d\sigma = 2\,\mathrm{artanh}\,\sigma \,.
\label{ssec}
\end{equation}
One verifies that $s$ agrees with the quadratic expansion in Eq.~(\ref{squad}), which supports our general method by way of example.

\subsection{Wave amplitude}

Next we calculate the wave amplitude ${\cal A}_0$ in geometrical optics. Unlike the rays of geodesics, waves experience the dimensionality of space, and so we need to use the full 3D metric here,
\begin{equation} 
ds^2 = n^2(dr^2 + r^2d\phi^2+ dz^2) \,,
\label{ds3d}
\end{equation}
and not only slices with $d\phi=0$. In terms of the Poincar\'{e} disk we have
\begin{equation} 
ds^2 = n_p^2(d\sigma^2 + \sigma^2d\theta^2) + n^2r^2d\phi^2 
\end{equation}
where $\theta$ denotes the angle on the disk. In $\{\sigma,\theta,\phi\}$ coordinates the equation of continuity for the amplitude ${\cal A}_0$ takes the form 
\begin{equation} 
\partial_\sigma \left(n\sigma r {\cal A}_0^2 \,\partial_\sigma s\right) = 0
\end{equation}
where we used the standard expression for the vector divergence in curved space \cite{GREEBook} and the convenient fact that $s$ depends only on $\sigma$. Additionally, $\partial_\sigma s = n_p$, and so
\begin{equation} 
n_p n \sigma r {\cal A}_0^2  =  f(\theta)
\label{a00}
\end{equation}
with some function $f$. Note that $f$ must be periodic, $f(\theta+2\pi)=f(\theta)$, because $\theta$ is an angle. Being a periodic function, we can express $f$ as a power series in
\begin{equation} 
e^{i\theta} = \frac{\xi+i\eta}{\sigma} \,.
\end{equation}
Furthermore, we know that for $r\rightarrow 0$ but $z\neq z_0$ the amplitude is finite (it only diverges at the point of emission where $z=z_0$). From this, and Eq.~(\ref{a00}), follows that $f\propto r$ for $r\rightarrow 0, z\neq z_0$. One sees from Eq.~(\ref{xieta}) that the only power series in $\exp(i\theta)$ with this property is proportional to $\eta/\sigma$. Hence we obtain
\begin{equation} 
{\cal A}_0^2 \propto \frac{\eta}{\sigma^2 n_p nr} = \frac{1-\sigma^2}{2\sigma^2 r}\, \frac{\sinh r \,\cos z \, \cos z_0}{\cosh r + \cos(z+z_0)} \,.
\end{equation}
As an additional bonus, we see that ${\cal A}_0$ is reciprocal:
\begin{equation} 
{\cal A}_0(z,z_0) = {\cal A}_0 (z_0,z) \,.
\end{equation}
The prefactor of ${\cal A}_0$ we obtain from the required asymptotics around the point of emission, Eq.~(\ref{a0}), which completes the calculation:
\begin{equation} 
{\cal A}_0 = - \frac{1}{4\pi \sigma} \sqrt{\frac{\eta}{n_p nr}} = - \frac{\cos z\, \cos z_0}{4\pi c_+ c_-} \sqrt{\frac{\sinh r}{r}} 
\label{a0sec}
\end{equation}
where the $c_\pm$ are defined in Eq.~(\ref{cpm}). One verifies again that this result agrees with Eq.~(\ref{ampex}) in quadratic expansion. 

\subsection{Geometric dispersion}

Finally, we calculate the amplitude $b$ for the geometric dispersion in the secant profile. We obtain from Eq.~(\ref{beta1}) with $R_z^z=-2$ and the amplitude ${\cal A}_0$ of Eq.~(\ref{a0sec}):
\begin{eqnarray} 
2\beta_1 &=& \frac{1}{n^2{\cal A}_0} \left(\frac{\partial_r r\, \partial_r {\cal A}_0}{r} + \frac{\partial_z n\, \partial_z {\cal A}_0}{n}\right) +1 \nonumber\\
&=& \frac{1}{4n^2} \left(1 + \frac{1}{r^2} - \frac{1}{\sinh^2 r}\right) \,.
\end{eqnarray}
One verifies that 
\begin{equation} 
(\partial_r s) (\partial_r \beta) + (\partial_z s) (\partial_z \beta) = n^2\beta_1 
\end{equation}
for the geodesic length $s$ given by Eq.~(\ref{sigma}-\ref{ssec}) and 
\begin{equation} 
\beta = \frac{c_+ c_-}{8r\sinh r} \left(r^2 + r\coth{r} -1\right) \,.
\end{equation}
According to Eq.~(\ref{beta1}), this $\beta$ is exactly the one needed for calculating the geometric dispersion $b={\cal A}_0 \beta$. We obtain from our result for ${\cal A}_0$, Eq.~(\ref{a0sec}), and Eq.~(\ref{sec}), the formula:
\begin{eqnarray} 
b &=& -\frac{1}{n_0 n}\sqrt{\frac{r}{\sinh r}}\,\frac{r^2 + r\coth{r} -1}{32\pi r^2} \,
\label{bsec}
\\
&=& - \frac{1}{24\pi n_0^2} \quad\mbox{for}\quad r \rightarrow 0 \,,
\end{eqnarray}
which agrees with our general result, Eq.~(\ref{bresult}), for exact geometries, Eq.~(\ref{exactgeo}). We also see that the amplitude $b$ described by Eq.~(\ref{bsec}) is finite and exponentially localized around the $z$--axis. There the mismatch between the 2D geometry of the geodesics, Eq.~(\ref{ds2d}), and the 3D geometry of the waves, Eq.~(\ref{ds3d}), is strongest, as all the radial planes go through the $z$--axis. In $2D$ the underlying geometry is maximally symmetric \cite{Maximal}. If this would also hold in 3D, no scattering could occur. However, the dimensional mismatch the wave experiences near the $z$--axis causes geometric dispersion and hence reflection inside the material that proliferates to multiple scattering \cite{LLRemark}, as the Casimir stress shows (Fig.~\ref{itay}).

\section{Conclusions}

We have derived a renormalisation procedure for the Lifshitz theory \cite{Lifshitz,Schwinger} of the Casimir force in inhomogeneous planar media with electric permittivity $\varepsilon$ and magnetic permeability $\mu$ as functions of Cartesian coordinate $z$ and frequency $\omega$. Our procedure amounts to the following recipe: subtract from the spectral stress density ${\cal W}$, defined in Eq.~(\ref{fourier}), the renormaliser
\begin{equation} 
{\cal W}_0 = -2w + \frac{n^2 (\partial_z Z)^2w^4 + (\partial_z n)^2Z^2 u^4}{4^2Z^2w^5}
\label{renormalisation}
\end{equation}
with $n$ being the refractive index $\sqrt{\varepsilon\mu}$, $Z$ the impedance $\sqrt{\mu/\varepsilon}$, and $w=\sqrt{\varepsilon\mu\kappa^2+u^2}$, while $\kappa$ denotes the wavenumber to the imaginary frequency $\omega=ic\kappa$ and $u$ the magnitude of the spatial Fourier component.

We have proved mathematically (Appendix A and B) that this simple procedure removes all infinities in the stress, apart from a logarithmic divergence that removes itself due to dispersion (the frequency dependance of $\varepsilon$ and $\mu$). More importantly, we have given a physical justification and a physical picture for the renormalisation procedure: the Casimir effect is brought about by quantum--fluctuating sources in dielectric media (Fig.~\ref{quantumnoise}b); our procedure removes the unphysical direct interaction of the sources with themselves (Fig.~\ref{pointsplit}b). For this, we have clarified the concept of the outgoing wave in inhomogenous media, using geometrical optics (Fig.~\ref{outgoing}) in quadratic expansion (Fig.~\ref{quadratic}). Our theory justifies both mathematically and physically a previous conjecture \cite{Philbin} that, as we have shown, amounts to the same procedure. The theory may become the starting point of predicting internal Casimir forces and their experimental consequences \cite{Paper2}.

Our regularising Green function is not reciprocal between emitter and receiver --- it distinguishes the emitter, because it depends on the quadratic expansion of $\varepsilon$ and $\mu$ around the emission point, whereas the equivalent Green function \cite{Philbin} is reciprocal, $g_{0}(z,z_0)=g_{0}(z_0,z)$. This feature suggests an important corollary: there is no trace anomaly for quantum electromagnetism in planar media. Trace anomalies have been discussed in connection with Casimir forces in curved space \cite{Wald}. There the renormalising Green function was non--reciprocal, which caused a violation of the energy--momentum conservation, unless a correcting term was introduced: the anomaly \cite{Wald}. The energy--momentum tensor of electromagnetic fields is traceless, but not the correction; hence the name trace anomaly. In the language of macroscopic electromagnetism, the conservation of the energy--momentum tensor in general relativity corresponds to the well-known relation \cite{LL8}:
\begin{equation} 
\bm{\nabla}\cdot\sigma + \frac{\bm{\nabla}\varepsilon}{2\varepsilon}\langle\bm{D}\cdot\bm{E}\rangle + \frac{\bm{\nabla}\mu}{2\mu}\langle\bm{B}\cdot\bm{H}\rangle = 0
\label{stressrelation}
\end{equation}
for the Maxwell stress $\sigma$ of fields with stationary average $\langle\rangle$. Renormalisation may introduce an additional pressure, the anomaly, but apparently not in planar media. 

Another interesting aspect of our theory is the role of dispersion. The principal singularities of the Casimir stress are caused by the evanescent waves around the source, not by waves of arbitrarily high frequencies that, in physical materials, cease to have an effect, as $\varepsilon\rightarrow\mu\rightarrow 1$ in the limit $\omega\rightarrow\infty$ for all real materials. The logarithmic singularity of the stress disappears due to dispersion; the divergence only remains for dispersionless models, apart from exceptional cases, such as the piece--wise homogenous materials where both Casimir's \cite{Casimir} and Lifshitz' \cite{Lifshitz} original theory had worked. The test case for Lifshitz' complicated theory of vacuum forces in realistic media was Casimir's simple formula \cite{Casimir} for the force between two perfect, dispersionless mirrors. Had the stress not converged in this exceptional case, Lifshitz' theory \cite{Lifshitz} would have hardly been trusted in the early days. Casimir theory came into existence thanks to a lucky exception.

\section*{Acknowledgements}

We thank 
Yael Avni,
Michael Berry,
Iver Brevik,
Yehonathan Drori,
Efi Efrati,
Mathias Fink,
Grzegorz Lach,
Mordehai Milgrom,
Kimball Milton,
Michael Moshe,
Sahar Sahebdivan,
Ephraim Shahmoon,
William Simpson,
Adiel Stern,
and Yana Zilberg
for stimulating discussions. 
Itay Griniasty is grateful to the Azrieli Foundation for the award of an Azrieli Fellowship. Our work was also supported by the European Research Council and the Israel Science Foundation, a research grant from Mr. and Mrs. Louis Rosenmayer and from Mr. and Mrs. James Nathan, and the Murray B. Koffler Professorial Chair.

\appendix

\section{Asymptotic expansion}

The divergence of the electromagnetic stress in the vacuum state appears in the high-frequency limit of the integrals
\begin{equation}
\sigma_{zz} = -\frac{\hbar c}{(2\pi)^2} \lim_{\Lambda\to\infty}\int_0^\Lambda \int_0^\Lambda {\cal W}\,u\,du\,d\kappa 
\label{stressa}
\end{equation}
with ${\cal W}$ given by Eq.~(\ref{fourier}). In order to analyse this limit we mentally replace $u\to u/q$ and $\kappa\to\kappa/q$ and consider the limit $q\rightarrow 0$. First we construct the Green functions as asymptotic series in $q$ using the Wronskian method. For this we need two appropriate solutions of the homogeneous wave equation,
\begin{equation}
\partial_z \frac{1}{\nu} \partial_z\, \widetilde{h} = \frac{u^2+n^2\kappa^2}{\nu q^2}\, \widetilde{h} \,,
\label{homodiff}
\end{equation}
and their Wronskian. We write $\widetilde{h}$ with the ansatz
\begin{equation}
\widetilde{h} =\exp\left(-\frac{1}{q}\sum_{m=0}^\infty q^m s_{m}\right)
\end{equation}	
and obtain
\begin{eqnarray}
\widetilde{h}^{-1}\,\partial_z \frac{1}{\nu}\partial_z \,\widetilde{h} &=& \frac{1}{q^2} \sum_{m=0}^\infty q^m \sum_{k=0}^m \frac{s_k' s_{m-k}'}{\nu} 
\nonumber\\
&& + \frac{1}{q} \sum_{m=0}^\infty q^m \left(-\frac{s''}{\nu}+\frac{\nu' s_m'}{\nu^2}\right)
\end{eqnarray}
where the primes denote derivatives with respect to $z$. The lowest--order contribution to this series is $q^{-2}s_0'^2/\nu$ which must compensate the right--hand side of Eq.~(\ref{homodiff}). Therefore we have
\begin{equation}
s_0'= \pm \sqrt{u^2 + n^2\kappa^2} \,.
\label{initials}
\end{equation}	
For the higher coefficients the right--hand side of Eq.~(\ref{homodiff}) is zero, and we get for $q^{m-2}$, $m>0$:
\begin{eqnarray}
0 &=& \sum_{k=0}^m s'_k s'_{m-k} - s''_{m-1} + \frac{\nu'}{\nu}\, s'_{m-1}
\nonumber \\
&=& 2s_0s'_m + \sum_{k=1}^{m-1} s'_k s'_{m-k} - s''_{m-1} + \frac{\nu'}{\nu}\, s'_{m-1}
\end{eqnarray}
and hence
\begin{equation}
s'_m= \frac{1}{2s'_0} \left(s''_{m-1} - \frac{\nu'}{\nu}\, s'_{m-1} - \sum_{k=1}^{m-1} s'_k s'_{m-k} \right) .
\label{recurrance}
\end{equation}	
This recurrence relation generates all the $s'_m$ from the initial $s'_0$ of Eq.~(\ref{initials}). The two sign choices for $s'_0$ define two linearly--independent homogeneous solutions $\widetilde{h}_\pm$. Note that the $s'_m$ with even $m$ changes sign if $s'_0$ does, whereas the $s'_m$ with odd $m$ are independent of the sign of $s'_m$. We thus write the two solutions as 
\begin{equation}
\widetilde{h}_\pm = e^{\pm s_\even}e^{s_\odd}
\label{htildepm}
\end{equation}	
where we define the even and odd series
\begin{equation}
s_\even\equiv\frac{1}{q}\sum_{m=0}^\infty q^{2m} s_{2n}\,,\quad
s_\odd\equiv\sum_{m=0}^\infty q^{2m} s_{2m+1}
\label{evenodd}
\end{equation}	
and adopt the positive sign for $s'_0$. The generalized Wronskian for the two solutions, defined as 
\begin{equation}
W(a) = \frac{1}{\nu} \left[ \widetilde{h}'_+(a)\,\widetilde{h}_-(a) - \widetilde{h}_+(a)\,\widetilde{h}'_-(a)\right] \,,
\label{wronski}
\end{equation}	
is a constant, $\partial_a W(a) =0$, as a consequence of Eq.~(\ref{homodiff}). The Green functions, the solutions of the inhomogeneous Eq.~(\ref{gdiff}), can be expressed in terms of the homogeneous solutions and the Wronskian as
\begin{equation}
g(z,z_0)= \frac{1}{W}
	\begin{cases}
		\widetilde{h}_+(z)\widetilde{h}_-(z_0)\,:\,  z>z_0\\
		\widetilde{h}_+(z_0)\widetilde{h}_-(z)\,:\,  z<z_0\,.
	\end{cases}
\label{wronskig}
\end{equation}	
Substituting Eqs.~(\ref{htildepm}) and (\ref{evenodd}) into Eq.~(\ref{wronski}) we may get an expression for the Wronskian --- we see that it is negative, but it is wiser to utilize the constancy of $W$ and write it as $W=-\sqrt{W(z)W(z_0)}$, which gives
\begin{equation}
W=-2\sqrt{\frac{s_\even'(z) s_\even'(z_0)}{\nu(z)\nu(z_0)}}\,e^{s_\odd(z)+s_\odd(z_0)} 
\end{equation}	
that we use in Eq.~(\ref{wronskig}) with Eq.~(\ref{htildepm}) for the $\widetilde{h}_\pm$. We thus obtain for the Green function the asymptotic formula
\begin{equation}
	g(z,z_0)=-\frac{1}{2}\sqrt{\frac{\nu(z)\nu(z_0)}
		{s_\even'(z) s_\even'(z_0)}}
	\exp\left (-\left | \int_{z_0}^{z}s_\even' dz
	\right |
	\right ).
\label{asygreen}
\end{equation}
We see that the Green function does only depend on the even series in Eq.~(\ref{evenodd}). Note also that the {\it ad hoc} renormaliser \cite{Philbin}, Eq.~(\ref{adhoc}), is the zeroth--order expansion in $q$ of the asymptotic Green function.

For working out the asymptotics of the stress, we use the asymptotic form of the Green function, Eq.~(\ref{asygreen}), with the recurrence relation of Eq.~(\ref{recurrance}) and the initial value of Eq.~(\ref{initials}) in the calculation of the stress, Eq.~(\ref{stressa}), with ${\cal W}$ given by Eq.~(\ref{fourier}). Note that in the integrals of Eq.~(\ref{stressa}) $du\to du/q$ and $d\kappa\to d\kappa/q$. After lengthy calculations we find three terms diverging with the cut--off $\Lambda=1/q$:

\begin{align}
	&\sigma_{zz} = \alpha_4\Lambda^4+\alpha_2\Lambda^2+\alpha \log\Lambda
	+ \text{ finite ,}
	\\
	&\alpha_4\Lambda^4 =
	\frac{\hbar c}{(2\pi)^2} \int^\Lambda \int^\Lambda 2w\,u\,du\,d\kappa \,,
	\label{l4}\\
	&\alpha_2\Lambda^2 =
	-\frac{\hbar c}{(2\pi)^2} \int^\Lambda \int^\Lambda \frac{n^2 Z'^2w^4 + n'^2Z^2 u^4}{4n^2Z^2w^5}\, u\,du\,d\kappa \,,
	\label{l2}\\
	&\alpha\log \Lambda =
	\frac{\hbar c}{(2\pi)^2}
	\int^\Lambda\int^\Lambda
	\frac{1}{16 w^{11} n^4 Z^4}\times
	\nonumber\\&
	\times\bigg [
	-20 u^8 Z^4 n'^4
	+u^2 w^6 n Z 
	\Big\{
	2 Z^2 
	\Big(
	5 n n'^2 Z''+19 n'^3 Z'
	\nonumber\\&
	\qquad
	+n^2 \left(n^{(3)} Z'-2 n'' Z''\right)
	+n n' \left(n Z^{(3)}-12 n'' Z'\right)
	\Big)
	\nonumber\\&
	\qquad
	+5 n^2 n' Z'^3
	+n Z Z' \left(n n'' Z'-10 n n' Z''+6 n'^2 Z'\right)
	\Big\}	
	\nonumber\\&
	\qquad
	+u^6 w^2 Z^3 n'^2 
	\left \{
	35 n n' Z'+9 Z \left(4 n'^2-n n''\right)
	\right\}
	\nonumber\\&
	\qquad
	+u^4 w^4 Z^3 
	\Big\{
	-5 n n' \left(-4 n n'' Z'+n n' Z''+15 n'^2 Z'\right)
	\nonumber\\&
	\qquad
	-2 Z \left(-n^2 n''^2+8 n'^4+n^2 n^{(3)} n'-5 n n'^2 n''\right)
	\Big\}
	\nonumber\\&
	\qquad
	+w^8 n^2 
	\Big\{
	3 n Z Z'^2 \left(n Z''-n' Z'\right)-2 n^2 Z'^4
	\nonumber\\&
	\qquad
	+2 Z^2 
	\left(
	n \left(n'' Z'^2+n Z''^2-n Z^{(3)} Z'\right)
	\right .
	\nonumber\\&
	\qquad
	-2 n'^2 Z'^2+n n' Z' Z''
	\Big)
	\Big\}
	\bigg ]\, u\,du\,d\kappa\,.		
\end{align}
The logarithmic divergence originates from integrations of the type
\begin{equation}
	\int^\Lambda\int^\Lambda 
	\frac{f(n^{(m)},Z^{(m)})}{w^{3}} \left (\frac{u}{w}\right )^l u\,du\,d\kappa \,,\quad
	\, l\geq0\,.
\end{equation}
Due to dispersion both impedance and refractive index tend to unity for high frequencies:
\begin{equation}
n \sim 1 + \frac{n_\infty}{\kappa} \,,\quad 
Z \sim 1 + \frac{Z_\infty}{\kappa} \,.
\end{equation}
All terms in \(\alpha\log\Lambda\) depend on derivatives of \(n\) or \(Z\), hence in dispersive media \(\alpha\log\Lambda\) behaves at high frequencies as
\begin{equation}
	\int^\Lambda\int^\Lambda 
	\frac{f(n_\infty^{(m)},Z_\infty^{(m)})}{\kappa w^{3}} \left (\frac{u}{w}\right )^l u\,du\,d\kappa\,,\quad
	\, l\geq0\,,
\end{equation}
which does not diverge.

If one considers a standard Lorentzian dispersion  \(n\sim 1+\kappa^{-2}\) for $\kappa\rightarrow\infty$, the term
\(\alpha_4\Lambda^4\) is unaffected, and \(\alpha_2\Lambda^2\) turns into a linear divergence; the infinity does remain, requiring theoretical aid in its removal.

\section{Stress of the outgoing wave}

It is straightforward to calculate the stress of the outgoing wave in real space. However, in planar media it is advantageous to calculate the stress in Fourier space with respect to the transverse coordinates $x$ and $y$. The divergences of the stress then appear through integration to infinity rather than in the limit ${\bm{r}}\to{\bm{r}}_0$, and are subtracted from the full Fourier--transformed stress. In this Appendix we calculate the required renormalizer in Fourier space. 

Our starting point is the Green function, Eq.~(\ref{ddef}), of the outgoing wave, Eq.~(\ref{d0}), in real space. In quadratic expansion, the amplitude ${\cal A}_0$ is given by Eq.~(\ref{ampex}) and the geodesic length $s$ by Eq.~(\ref{squad}). Note that the $r^2$ term in $s$ would make the Fourier--transform of the Green function divergent if we take it literally for large $r$ and not regard it as what it is: an expansion around the point of emission. We thus expand this contribution in $\exp(-\kappa s)$, and get
\begin{equation}
g_0 \sim \nu\nu_0{\cal A}_0 \left(1+\kappa\rho\,\frac{n_0'^2 r^2}{24n_0}\right) e^{-\kappa \rho \chi} \sim \sum_{m=1}^3 g_m
\end{equation}
with the abbreviations
\begin{eqnarray} 
\chi &=& n_0 + \frac{n_0'}{2}(z-z_0) + \frac{n_0''}{6} (z-z_0)^2 \,,\nonumber\\
g_1 &=& -\frac{\sqrt{\nu \nu_0}}{4\pi\rho}\, e^{-\kappa\rho\chi} \,,\nonumber\\
g_2 &=& -\frac{\sqrt{\nu \nu_0}}{4\pi}\,\rho\,\frac{n_0^2 R}{48}\, e^{-\kappa\rho\chi} \,,\nonumber\\
g_3 &=& -\frac{\sqrt{\nu \nu_0}}{4\pi}\,\kappa\,\frac{n_0'^2 r^2}{24n_0}\,e^{-\kappa\rho\chi}
\end{eqnarray}
where $\rho=\sqrt{r^2+(z-z_0)^2}$ and $R$ is the curvature scalar given by Eq.~(\ref{curvature}); primes denote differentiations. The first term $g_1$ of the Green function we Fourier--transform, 
\begin{equation}
\widetilde{g} = 2\pi \int_0^\infty g J_0(u r)\, r\,dr \,,
\label{besselfourier}
\end{equation}
and obtain from Eq.~2.12.10.10 of Ref.~\cite{Prudnikov}
\begin{equation}
\widetilde{g}_1 = -\frac{\sqrt{\nu \nu_0}}{2W}\,\exp\left(-W\left|z-z_0\right|\right) 
\label{g1tilde}
\end{equation}
with 
\begin{equation}
W= \sqrt{u^2+\chi^2\kappa^2} \,.
\label{wchi}
\end{equation}
We then calculate its spectral stress density ${\cal W}_1$ according to Eq.~(\ref{fourier}) and get
\begin{equation}
{\cal W}_1 = -2w + \frac{Z'^2}{4wZ^2} + \frac{w^4n^{-2}-3u^2\kappa^2}{4w^5}\,n'^2 - \frac{\kappa^2 n n''}{6w^3}
\label{ww1}
\end{equation}	
where $w$ abbreviates $w=\sqrt{u^2+n^2\kappa^2}$ and $Z$ denotes the impedance $Z=\sqrt{\mu/\varepsilon}$, as in Eqs.~(\ref{nZ}) and (\ref{w2}). The first term in Eq.~(\ref{ww1}) gives the $\Lambda^4$ divergence of the stress, while the other terms make contributions to the $\Lambda^2$ divergence. 

The other two contributions ${\cal W}_2$ and ${\cal W}_3$ to the spectral stress density we calculate in real space first and then Fourier--transform them.  We get for $g_2$:
\begin{equation}
\left.\sum_\mathrm{p} \frac{1}{\nu}\left(w^2-\partial_z\partial_{z_0}\right)g_2\right|_{z_0\to z} \!\!\!\!\!\!\!\!\!\! \sim -\frac{\kappa(n'^2-2nn'')}{12\pi n} \, e^{-\kappa n r}
\end{equation}	
where we used $w^2=n^2\kappa^2 - r^{-1}\partial_r r \partial_r$ (expressed in cylindrical coordinates) and took the limit $r\to 0$ in the prefactor of $e^{-\kappa n r}$. We need to keep this exponential factor for getting a finite Fourier transform. We perform the Fourier--transformation of the result according to the Bessel--Fourier formula of Eq.~(\ref{besselfourier}). For this we use Eq.~2.12.8.4 of Ref.~\cite{Prudnikov}
\begin{equation}
\int_0^\infty  \,e^{-\kappa n r} r^{\alpha -1}J_0( u r)\,dr=w^{-\alpha}\,\Gamma(\alpha) P_{\alpha-1}\left(\frac{\kappa n}{w} \right)
\label{prud}
\end{equation}	
for $\alpha>0$ where $P_n$ are the Legendre polynomials \cite{Erdelyi}, $\Gamma$ the Gamma function \cite{Erdelyi} and $w=\sqrt{u^2+n^2\kappa^2}$. In particular, we have for $\alpha=2$:
\begin{equation}
\int_0^\infty  \,e^{-\kappa n r} J_0( u r)\,r\,dr=\frac{\kappa n}{w^3} \,.
\end{equation}	
In this way we obtain
\begin{equation}
{\cal W}_2 = \frac{\kappa^2(2n n'' - n'^2)}{12w^3} \,,
\end{equation}	
which contributes to the $\Lambda^2$ divergence in the stress.

For calculating the spectral stress density of $g_3$ we follow the same procedure, except that we take into account also the linear term in the prefactor of $e^{-\kappa n r}$:
\begin{equation}
\left.\sum_\mathrm{p} \frac{1}{\nu}\left(w^2-\partial_z\partial_{z_0}\right)g_3\right|_{z_0\to z} \!\!\!\!\!\!\!\!\!\! \sim \frac{n'^2(\kappa -\kappa^2 n r)}{12\pi n} \, e^{-\kappa n r}
\end{equation}	
because this term is proportional to $\kappa^2$ and hence also contributes to the $\Lambda^2$ divergence if we Bessel--Fourier transform it with 
\begin{equation}
\int_0^\infty  \,e^{-\kappa n r} J_0( u r)\,r^2\,dr=\frac{3\kappa^2n^2-w^2}{w^5} 
\end{equation}	
according to Eq.~(\ref{prud}). We obtain
\begin{equation}
{\cal W}_3 = \frac{\kappa^2n'^2(3u^2-w^2)}{6w^3} \,.
\end{equation}	
Finally, combining all three contributions,
\begin{equation}
{\cal W}_0 = {\cal W}_1 + {\cal W}_2 + {\cal W}_3 \,,
\end{equation}	
we arrive at Eq.~(\ref{renormalisation}) that describes in Fourier space the real--space divergence of the stress due to the outgoing wave. Note that there is no non--zero finite contribution that needs to be accounted for (in the order of limits required for the Fourier method).

\section{Linear expansion}

The principal divergences of the stress, Eqs.~(\ref{l4}) and (\ref{l2}), do only depend on the values and first derivatives of the refractive index $n$ and impedance $Z$. One might therefore be inclined to expand the outgoing wave up to linear order only. In this Appendix we show, however, that this linear expansion produces a second derivative of the refractive index in the stress (and disagrees with the divergencies). 

In linear expansion, the Green function, Eq.~(\ref{ddef}), of the outgoing wave, Eq.~(\ref{d0}), is simply:
\begin{equation}
g_0 \sim -\frac{\sqrt{\nu \nu_0}}{4\pi\rho}\, e^{-\kappa\rho\chi} \,,\quad \chi = n_0 + \frac{n_0'}{2}(z-z_0) 
\end{equation}
where we considered both the geodesic distance, Eq.~(\ref{squad}), and the amplitude, Eq.~(\ref{ampex}), in linear expansion. We follow the same procedure as in Appendix B and Fourier--transform $g_0$ with the same result as in Eqs.~(\ref{g1tilde}) and (\ref{wchi}). However, as $\chi$ is truncated to linear order, we obtain a different result for the spectral stress density:
\begin{equation}
{\cal W}_0 = -2w + \frac{Z'^2}{4wZ^2} + \frac{w^4n^{-2}-3u^2\kappa^2}{4w^5}\,n'^2 - \frac{\kappa^2 n n''}{2w^3} \,.
\label{cresult}
\end{equation}	
Moreover, the curvature term from the quadratic expansion of the amplitude, Eq.~(\ref{ampex}) is missing that compensates for the second derivative in the stress, and so is the contribution from the $r^2$--term of the geodesic length, Eq.~(\ref{squad}). The result, Eq.~(\ref{cresult}), takes care of the $\Lambda^4$ divergence of the stress, Eq.~(\ref{l4}), but fails to compensate for the $\Lambda^2$ divergence, Eq.~(\ref{l2}).

Alternatively, one could linearly expand $\varepsilon$ and $\mu$ only, while using a quadratic expansion for the outgoing wave; this does not work either. In any case, a linear expansion of $\varepsilon$ and $\mu$ is not sufficient to capture the divergencies of the Casimir stress.



\begin{thebibliography}{99}
\bibitem{Piecewise}
One sees from Eq.~(\ref{stressrelation}) that the force density $\bm{\nabla}\cdot\sigma$ vanishes in homogeneous regions where $\bm{\nabla}\varepsilon=\bm{\nabla}\mu=0$.
\end{thebibliography}
\end{document}